\documentclass[fleqn,usenatbib,twocolumn]{mnras}

\usepackage{newtxtext,newtxmath}

\usepackage[utf8]{inputenc}
\usepackage[T1]{fontenc}
\usepackage{ae,aecompl}
\usepackage{graphicx}	
\usepackage{amsmath}	
\usepackage{bm}
\usepackage{color}
\usepackage{hyperref}
\usepackage{afterpage}
\usepackage{lipsum}
\usepackage{url}

\newcommand{\br}{\mathbf{r}}
\newcommand{\bx}{\mathbf{x}}
\newcommand{\bu}{\mathbf{u}}

\newcommand{\bnr}{\mathbf{\nabla}_\mathbf{r}}
\newcommand{\bnx}{\mathbf{\nabla}_\mathbf{x}}
\newcommand{\erfc}{{\rm erfc}}
\newcommand{\xs}{x_{\rm s}}

\newcommand{\bp}{\mathbf{p}}
\newcommand{\hMpci}{h~{\rm Mpc}^{-1}}
\newcommand{\hiMpc}{h^{-1}{\rm Mpc}}
\newcommand{\bk}{\mathbf{k}}

\newcommand{\by}{\mathbf{y}}
\newcommand{\bs}{\mathbf{s}}
\newcommand{\bff}{\mathbf{f}}
\newcommand{\SM}[1]{\textcolor{black}{#1}}
\newcommand{\mtrv}[1]{\textcolor{black}{#1}}
\newcommand{\tnrv}[1]{\textcolor{black}{#1}}

\newcommand{\smrv}[1]{\textcolor{black}{#1}}

\title{Anisotropic separate universe simulations}
\author[S. Masaki, T. Nishimichi \& M. Takada]{Shogo Masaki$^{1,2}$
\thanks{shogo.masaki@gmail.com}
, Takahiro Nishimichi$^{3,4}$ and  
Masahiro Takada$^{4}$\\
$^{1}$Department of Mechanical Engineering, National Institute of Technology, Suzuka College, Suzuka, Mie 510-0294, Japan\\
$^{2}$Department of Physics, Nagoya University, Nagoya, Aichi 464-8601, Japan\\
$^{3}$Center for Gravitational Physics, Yukawa Institute for Theoretical Physics, Kyoto University,
Kyoto 606-8502, Japan\\
$^{4}$Kavli Institute for the Physics and Mathematics of the Universe (WPI),
The University of Tokyo Institutes for Advanced Study (UTIAS),\\
The University of Tokyo, Kashiwa, Chiba 277-8583, Japan}
\date{\today}

\begin{document}
\setlength{\mathindent}{0pt}
\label{firstpage}
\pagerange{\pageref{firstpage}--\pageref{lastpage}}
\maketitle

\begin{abstract}
The 
long-wavelength
coherent overdensity and tidal force, which are not direct observables for a finite-volume survey, affect time evolution of cosmic structure formation and therefore clustering observables through the mode coupling.
In this paper we develop an ``anisotropic'' separate universe (SU) simulation technique to simulate large-scale structure formation taking into account the effect of large-scale tidal force into the anisotropic expansion of local background. We modify the TreePM $N$-body simulation code to implement the anisotropic SU simulations, and then study the ``response'' function of matter power spectrum that describes how the matter power spectrum responds to the large-scale tidal effect as a function of wavenumber and redshift for a given global cosmology. We test and validate the SU simulation results from the comparison with the perturbation theory predictions and the results from high-resolution PM simulation\smrv{s}. We find that the response function displays characteristic scale dependences over the range of scales down to nonlinear scales, up to $k\simeq 6~h~{\rm Mpc}^{-1}$.
\end{abstract}
\begin{keywords}
large-scale structure of Universe -- cosmology: theory
\end{keywords}

\section{Introduction}
\label{sec:intro}

Large-scale structure of the universe is now recognized as a powerful probe of cosmology as it enables us to constrain properties 
of primordial fluctuations, explore the nature of dark matter and dark energy, and obtain precise measurements of cosmological parameters. For this reason, there are various ongoing and planned wide-area galaxy surveys such as the Subaru Hyper Suprime-Cam survey 
\citep{Aihara18}, the Subaru Prime Focus Spectrograph (PFS) survey \citep{2014PASJ...66R...1T}, 
the ESA Euclid mission \citep{laureijs2011}, the Legacy Survey of Space and Time\footnote{\url{https://www.lsst.org}},  
and the NASA WFIRST mission \citep{Spergel15}.

To attain the full potential of such wide-area galaxy surveys, there are both observational and theoretical challenges. On observation side we need high-precision characterization and measurements of the statistical properties of large-scale structure. On theory side we need accurate theoretical templates of the clustering observables to compare with the high-precision measurements at scales down to the nonlinear regime. However, there are unavoidable uncertainties in the large-scale structure cosmology. Such an example is the uncertainties arising from 
``super-sample'' modes \citep{TakadaHu13}
that are the effects of fluctuations with wavelengths comparable with or greater than a size of survey volume. 
Super-sample modes are not direct observables for a finite-volume survey, but  affect the time evolution of sub-survey modes and therefore 
clustering observables through the nonlinear mode coupling \citep{2006MNRAS.371.1188H,sato09,TakadaHu13}. The super-sample effects are also not easy to theoretically study, because $N$-body simulations, which are standard tools to study nonlinear structure formation, usually employ the periodic boundary conditions and ignore the effects of super-box modes  \citep{Sirko05,2011ApJS..194...46G}. 

There are two super-sample effects that are both related to the Hessian tensor of the long-wavelength gravitational potential and therefore of equal importance. 
The first is the coherent density contrast, and the effects on various clustering observables have been well studied
using the perturbation theory and $N$-body simulations 
\citep{Sirko05,2009MNRAS.395.2065T,2011JCAP...10..031B,sherwin12,TakadaHu13,2013MNRAS.429..344K,Li14,mohammed:2014lr,Wagner15a,Wagner15b,2016JCAP...09..007B,2018JCAP...06..015B,Takahashi19,Barreira19}. In particular, the effects can be absorbed into a change of cosmological parameters, especially the spatial curvature, in an isotropic Friedmann-Robertson-Walker (FRW) background. Hence we can run $N$-body simulations in the modified FRW background to fully study the super-sample effects over all scales down to the deeply nonlinear regime
-- so-called separate universe (SU) simulation technique \citep{Li14,Wagner15a,2016JCAP...09..007B}.
The SU simulations allow for an accurate calibration of the super-sample effects on desired observables, because paired SU simulations using the same seeds of initial conditions significantly reduce the sample variance errors in simulations. 

Another super-sample effect arises from the large-scale tidal force. The tidal effect causes an apparent anisotropic clustering in the large-scale structure depending on the degree of alignments between the wavevector and the tidal tensor in a given survey realization 
\citep{2012PhRvD..86h3513S,2015JCAP...10..059D,2017JCAP...02..025I,akitsu17,akitsu18,2018JCAP...02..022L,Akitsu19}. Most of the previous studies were based on the perturbation theory. 
One difficulty for a simulation based study is the large-scale tidal force cannot, by definition, be absorbed by a modification of an isotropic FRW background cosmology. Instead one needs to consider an ``anisotropic'' expansion to include the tidal effect into the 
local background, which we call an anisotropic SU simulation approach. \citet{schmidt18} made the first attempt to develop the anisotropic SU simulation technique, using the PM method. 

Hence the purpose of this paper is to develop the anisotropic SU simulation technique by modifying the TreePM algorithm in the publicly-available {\sc Gadget-2} code \citep{gadget2}. 
The TreePM\SM{, in which the tree force complements the poor accuracy of the PM force near and below the mesh size,} is one of the most powerful numerical method to simulate nonlinear structure formation including formation and properties of halos where galaxies and galaxy clusters form. Then we use the SU simulations to study the ``response'' function of matter power spectrum that describes how the large-scale tidal force affects the matter power spectrum as a function of wavenumber and redshift for a given global cosmology. 

The rest of this paper is structured as follows.
Section~\ref{sec:form} presents a formulation for anisotropic 
SU
simulations based on the TreePM method. 
After deriving the response function of the matter power spectrum to the large-scale tidal force in Section~\ref{sec:GK},
we study the response function using the SU simulations in Sections~\ref{sec:sim} and \ref{sec:res}, where we also give a validation of the SU simulations from the comparison with the perturbation theory prediction and with the high-resolution PM simulations.
Section~\ref{sec:conc} is devoted to conclusion and discussion. Throughout this paper we consider the standard $\Lambda$ and cold dark matter dominated cosmology with adiabatic Gaussian initial conditions
($\Lambda$CDM).

\section{Algorithm of separate universe simulation including large-scale tidal effect}
\label{sec:form}

\begin{figure*}
\begin{center}
\includegraphics[width=0.98\linewidth]{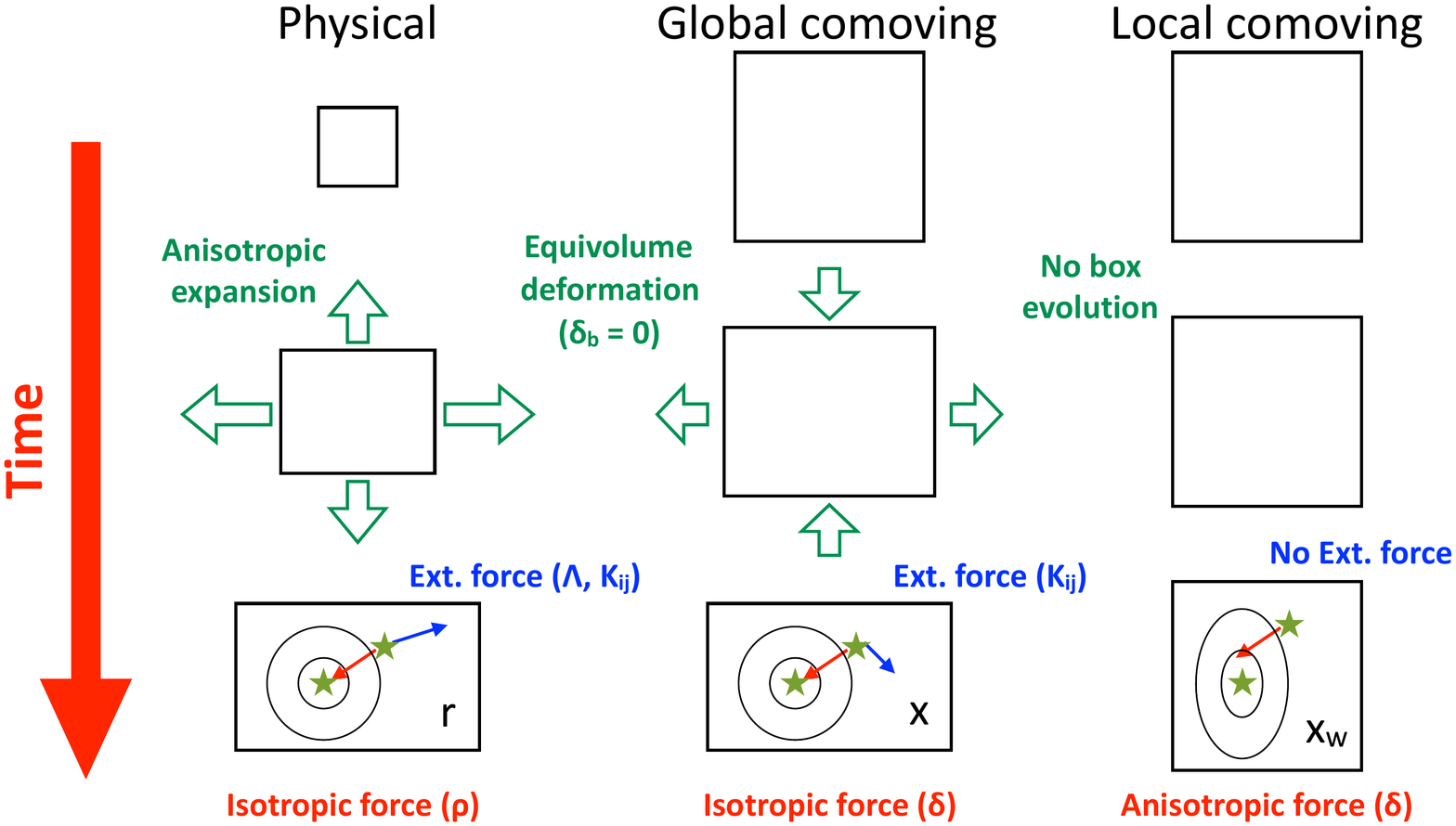}
\caption{A schematic illustration of structure formation in a finite volume region with the large-scale tidal field in 
different coordinate systems. \mtrv{The boundary box denotes
the Lagrangian volume at the initial epoch (upper panels), and we show}
how the 
volume in different coordinates evolves with time. The circular lines in the bottom panels depict 
\mtrv{iso-amplitude contours of the }
gravitational potential that is induced by a star-symbol particle at the center of the circle (ellipsoid). Then we consider the gravitational force acting on a test particle denoted by another star symbol. 
{\it Left}: The physical coordinate. The total gravitational force is given by a sum of the two contributions; the external gravitational force that arises from the spatially-homogeneous parts (the cosmological constant and the super-survey tidal tensor $K_{ij}(t)$), denoted by the blue arrow, and the gravitational force from the particle at the center, denoted by the red arrow. {\it Middle}: The standard comoving coordinate \SM{(the global comoving coordinate)}. The gravitational force corresponding to the isotropic expansion 
due to the mean matter density and the cosmological constant can be absorbed into the scale factor, $a(t)$, whose time evolution is given by the Friedmann equation. Then the gravitational force due to the super-survey tidal tensor needs to be given as an external force (blue arrow). The force from the mass overdensity with respect to the mean density ($\delta$) due to 
the particle at the center is denoted by the red arrow. Here we consider the region with the same volume as that in the physical coordinate at a target redshift (bottom panels).
{\it Right}: The comoving coordinate in the SU simulation \SM{(the local comoving coordinate)}. Now the force due to the super-survey tidal tensor ($K_{ij}$) is further absorbed into the comoving scale factor, $a_{Wi}(t)$. In this case the background has an anisotropic expansion. The gravitational force from the particle at the center becomes anisotropic\SM{: the force direction deviates from the direction 
\mtrv{connecting}
the two particles}. The simulation volume in the SU comoving coordinate appears to stay the same. 
We have to design the SU simulation so that 
the time evolution of self-gravitating system should be equivalent to those in the physical or global comoving coordinates. 
}
\label{fig:coord}
\end{center}
\end{figure*}

\mtrv{The purpose of this paper is to modify the publicly-available $N$-body code, {\sc Gadget-2} \citep{gadget, gadget2}, to include the effect of super-box 
tidal force on large-scale structure formation simulated in an $N$-body simulation of a finite volume -- so-called separate universe 
(SU) simulation. In this section, we}
present 
a formulation needed for the modification \mtrv{of the $N$-body code}. 
\mtrv{We first briefly review the formulation for a standard $N$-body code, i.e. $N$-body simulations in an isotropic, homogeneous Friedmann-Robertson-Walker (FRW) background, and then present a formulation for the $N$-body code in SU simulations we employ in this paper. Our formulation is based on the concept illustrated in Figure~\ref{fig:coord}.}

\subsection{Preliminaries: Newtonian $N$-body simulation equations in an \mtrv{isotropic} FRW background}
We begin by briefly reviewing basic equations used in standard $N$-body simulations in an isotropic, homogeneous FRW background.
We also define notations of quantities used in this paper.
To consider structure formation in an expanding FRW background, it is quite convenient to consider the gravitational evolution in the {\it comoving} coordinate, where an isotropic, homogeneous FRW expansion is solved separately, and the background equations are 
subtracted from a set of dynamical equations that govern structure formation including time evolution of density fluctuations.  
The comoving coordinate, denoted by $\bx$, is related to the physical 
coordinate, denoted by $\br$, via
%
\begin{align}
\br = a(t)\bx,\label{eq:comoving_iso}
\end{align}
where $a(t)$ is the scale factor that describes an expansion history of the FRW background. The time evolution of $a(t)$ 
in a matter dominated era is governed by the Einstein equation, the so-called Friedmann equation:
\begin{align}
\frac{\ddot{a}}{a} = -\frac{4\pi G\bar{\rho}(t)}{3}+\frac{\Lambda}{3},
\label{eq:Friedmann}
\end{align}
where the dot notation $\dot{\hspace{0.5em}}$ denotes the time derivative, 
$\bar{\rho}(t)$ is the mean matter density, given by $\bar{\rho}(t)=\bar{\rho}_0/a(t)^3$, $\bar{\rho}_0$ is the present-day 
matter density, and $\Lambda$ is the cosmological constant. The cosmological parameters are often used to specify 
$\bar{\rho}_0$ and $\Lambda$ as $\bar{\rho}_{\rm 0}=\Omega_{\rm m0}\bar{\rho}_{\rm cr0}$ and 
\tnrv{$\Lambda=8\pi G\Omega_{\Lambda}\bar{\rho}_{\rm cr0}$}, where $\bar{\rho}_{\rm cr0}$ is the critical density, defined as $\rho_{\rm cr0}\equiv 3H_0^2/8\pi G\simeq$
$1.88\times 10^{-29} h^2 $g cm$^{-3}$, for the convention of $c=1$ for 
the speed of light. \smrv{Throughout this paper we consider a flat-geometry universe: $\Omega_{\rm m0}+\Omega_\Lambda=1$.}

When the matter distribution has initial perturbations as predicted in the standard structure formation scenario, the spatial inhomogeneities grow due to the gravitational instability \citep{peebles1980,2003moco.book.....D}. Following the convention in
\citet{gadget}, we use
a normalized peculiar velocity, corresponding to the canonical momentum:
%
\begin{align}
\bu 
\equiv 
a^2\dot{\bx}.\label{eq:velocity_u_iso}
\end{align}
Given this definition,
we can introduce the effective Newtonian gravitational potential including the contribution of the cosmological constant, via the Poisson equation in the physical coordinate:
%
\begin{align}
\bnr^2 \phi(\br; t) = 4\pi G \rho(\br; t) - \Lambda,\label{eq:Poisson_iso}
\end{align}
where $\rho(\br; t)$ is the total matter density field and can be expressed in terms of the mean matter density ($\bar{\rho}$) and 
the matter density fluctuation field 
$\delta(\br;t)$
as 
$\rho(\br;t)\equiv \bar{\rho}(t)[1+\delta(\br;t)]$. \mtrv{The equation of motion for a test particle is given by}
%
\begin{align}
\ddot{\br} = -\bnr \phi = -\frac{1}{a}\bnx \phi.\label{eq:force_iso}
\end{align}
On the other hand,
the second \mtrv{time} derivative of Eq.~(\ref{eq:comoving_iso}) \mtrv{yields}
%
\begin{align}
\ddot{\br} = \ddot{a}\bx + 2\dot{a}\dot{\bx} + a\ddot{\bx}.
\end{align}
Using Eqs.~(\ref{eq:velocity_u_iso}) and (\ref{eq:force_iso}), we can rewrite the equation of motion 
as
%
\begin{align}
\dot{\bu} &= -\bnx\phi-a\ddot{a}x
\equiv 
-\frac{1}{a}\bnx\Phi.
\label{eq:peculiar_acceleration}
\end{align}
In the last equality on the r.h.s., we introduced the ``peculiar'' gravitational potential defined by
%
\begin{align}
\Phi \equiv  a \left[\phi + \frac{1}{2}a\ddot{a}x^2\right]. 
\end{align}
Using Eqs.~(\ref{eq:Friedmann}) and (\ref{eq:Poisson_iso}), we can find that the peculiar gravitational potential obeys the 
Poisson equation in the comoving coordinate:
\begin{align}
\bnx^2
\Phi(\bx; t) = 4\pi G \bar{\rho}_0\delta(\bx;t).
\label{eq:peculiar_Poisson_iso}
\end{align}
The use of the peculiar gravitational potential is convenient, because it arises from the density fluctuation field which is a spatially varying field, $\delta(\bx, t)$. Even if the gravitational force in the physical coordinate has source terms arising from 
the mean matter density and the cosmological constant (Eq.~\ref{eq:Poisson_iso}), the force contributions can be absorbed into the effect on the scale factor via the Friedmann equation. 
The development of SU simulation for the large-scale tidal effect is based on the similar concept, as we discuss below (see Figure~\ref{fig:coord}).

\mtrv{To perform a fast computation of the peculiar gravitational force and increase the dynamic ranges, 
we approximate the gravity force by a sum of the two parts that are computed based on the particle-mesh (PM) method and the tree algorithm -- the so-called TreePM code \SM{\citep{Xu95,2000ApJS..128..561B,2002JApA...23..185B}},
\begin{align}
\Phi(\bx;t)\simeq \Phi^{\rm PM}(\bx;t) + \Phi^{\rm T}(\bx;t).
\end{align}
Here the PM part computes the gravitational force on large scales treating a system of $N$-body particles as a coarse-grained fluid, while the tree part computes the direct gravitational force between $N$-body particles
on small scales. The division is to some extent arbitrary, and the parameters to determine the division need to be carefully tested and calibrated. When we include the effect of large-scale tidal force, we need to carefully study \smrv{whether the division in a SU simulation causes
an artifact,}
which is one of the main purposes of this paper.}

For the PM force calculation, it is \mtrv{useful to use the fast Fourier transform (FFT) method that allows for a fast computation of the gravitational field if grid-based data is given. From Eq.~(\ref{eq:peculiar_Poisson_iso}), the Fourier transform of the gravitational field is given by}
%
\begin{align}
\Tilde{\Phi}^{\rm PM}_{\bk}=-\frac{4\pi G \bar{\rho}_0 \Tilde{\delta}_{\bk}}{k^2}
=-4\pi G \bar{\rho}_0 \Tilde{\delta}_{\bk}G(k),
\end{align}
where $k^2\equiv \sum_i (k_i)^2$,
$G(k)\equiv 1/k^2$ is the Green function and $\Tilde{\delta}_{\bk}$ is the Fourier transformed density fluctuation. 
Fourier-transforming $\Phi^{\rm PM}_{\bk}$ back to real space gives the gravitational field \tnrv{due to the density field smoothed over the Fourier grids.} 
Here the number of FFT grids determines the resolution of the FFT based gravitational force for a given simulation box size; if we employ $N_{\rm grid}$ grids, the PM gravitational force is accurate on 
lager scales, $x\gg L_{\rm box}/N_{\rm grid}^{1/3}$ or 
$k\ll N_{\rm grid}^{1/3}k_{\rm f}$, where $k_{\rm f}$ is the fundamental Fourier mode, $k_{\rm f}=2\pi/L_{\rm box}$ ($L_{\rm box}$ is a size of $N$-body simulation box on a side). For the FFT computation we need to define the grid-based data from the distribution of $N$-body particles, and we adopt the Cloud in Cell (CIC) interpolation scheme 
in this paper. Consistently, the force interpolation from grid points is done with the same CIC algorithm. Following the original implementation of \textsc{Gadget-2}, we devide the potential field by the square of the CIC kernel in Fourier space to account for the two interpolation steps together with the multiplication of the Green function described below. We, however, omit the window function from the equations in what follows for notational simplicity.
What is important in this process is that the FFT method implicitly assumes the periodic boundary conditions of the data. While this is advantageous in the sense that it automatically takes into account the force from infinite mirror images of the simulation box,
the method ignores the effect of super-box modes with $k<k_{\rm f}$ by construction. 
Another advantage of the PM method is that, as long as we can increase $N_{\rm grid}$ or equivalently the Fourier resolution for a fixed simulation box size, the gravity is very accurately simulated at scales greater than the FFT limit. Hence we can use the PM results with high resolution to test the TreePM method for the SU simulations.
Finally, to leave gravity in short ranges to the Tree algorithm, the PM potential in the Fourier space needs to be smoothly truncated around FFT grid scales, which is done  
by the split factor:
\begin{align}
    \Tilde{\Phi}^{\rm PM}_{\bk}
    \rightarrow
    \Tilde{\Phi}^{\rm PM}_{\bk}\exp\left(-k^2~\xs^2\right),
\end{align}
where $\xs$ is the comoving split scale.
For our computation, we set $\xs=4.5 L_{\rm box}/N_{\rm grid}^{1/3}$.

On the other hand,
the gravitational potential in short ranges is 
computed based on the Tree algorithm. 
The gravitational potential at the position of the $n$-th $N$-body particle is computed from a direction summation of the $1/r$-force between $N$-body particles:
\begin{align}
\Phi^{\rm T}(\bx_n)
&=-Gm\sum_{n'; n'\ne n}\frac{1}{|\bx_n-\bx_{n'}|}\erfc\left(\frac{|\bx_n-\bx_{n'}|}{2\xs}\right),
\label{eq:iso_tree_pot}
\end{align}
where the summation runs over all the $n'$-th particles except the $n$-th particle (i.e.
$n'\ne n$), 
$m$ is the particle mass,
and ${\rm erfc}(x)$ is the complementary error function. The factor 
with $\erfc$ is the split factor for the Tree force;
the gravitational force contribution from the $n'$-th particle in large distances $|\bx_n-\bx_{n'}|\gg \xs$ is suppressed because 
${\rm erfc}(x)\simeq 0$ for $x\gg 1$.
Note that the contribution from the mean density should be subtracted from this expression because it is already absorbed into the coordinate transformation. However, with the periodic boundary condition employed in the simulations, this does not affect the motion of simulation particles at all.
In an isotropic universe background, 
the resultant Tree gravitational acceleration is
\begin{align}
&-\bnx \Phi^{\rm T}(\bx_n)
=-Gm\sum_{n';n'\neq n}\frac{\bx_n-\bx_{n'}}{|\bx_n-\bx_{n'}|^3}\nonumber\\
&\times\left[
\erfc\left(\frac{|\bx_n-\bx_{n'}|}{2\xs}\right)
+\frac{|\bx_n-\bx_{n'}|}{\xs\sqrt{\pi}}\exp\left(-\frac{|\bx_n-\bx_{n'}|^2}{4\xs^2}\right)
\right] . 
\label{eq:tree_iso_com}
\end{align}
The factor of ($-1/|\bx_n-\bx_{n'}|$) in Eq.~(\ref{eq:iso_tree_pot}) is replaced with a kernel incorporating the softening length to avoid gravity divergence \citep{gadget}.
In {\sc Gadget-2}, the gravitational potential for a group of particles is expanded in multipole series, and the gravitational force is calculated only with the monopole term \citep{gadget2}.
Together with the split factor, the final Tree acceleration form is
\begin{align}
    -\nabla_\bx \Phi^{\rm T}(\bx_n)
    &=G\sum_{\rm group} M_{\rm group}g_1(y_{\rm group})\by_{\rm group}\nonumber\\
    &\times\left[
    \erfc\left(\frac{y_{\rm group}}{2\xs}\right)
    +\frac{y_{\rm group}}{\xs\sqrt{\pi}}\exp\left(-\frac{y_{\rm group}^2}{4\xs^2}\right)
    \right],
\end{align}
where $\by_{\rm group}=\bx_n-\bs_{\rm group}$, $\bs_{\rm group}$ is the center-of-mass of the particle group with the total mass $M_{\rm group}$ and $g_1$ is the monopole term of the potential expanded in the multipole series \citep{gadget}.
In the summation above, since the Tree force is confined to small scales, we need to consider only the nearest image out of the infinite periodic mirrors.

The ``peculiar'' Hamiltonian for a system of $N$-body particles is 
\begin{align}
H=\sum_n\frac{\bp^2_n}{2 m a^2}
+\frac{1}{2}\sum_{n}\frac{m\Phi(\bx_n)}{a},
\end{align}
where the first term is the kinetic energy \tnrv{with the canonical momentum defined by $\bp=m\bu$} and the second term is the gravitational potential arising from the inhomogeneous distribution of $N$-body particles (that is, the homogeneous parts such as $\bar{\rho}$ and $\Lambda$ are 
implicitly included only in the time evolution of the scale factor $a$
in this Hamiltonian).
{\sc Gadget-2} employs the time-evolution operators for the kinetic and the potential parts of the Hamiltonian \citep{quinn97,gadget2}:
\begin{align}
D(\Delta t):&~
\begin{cases}
    \displaystyle \bp_n \rightarrow \bp_n \\
    \displaystyle \bx_n 
    \rightarrow 
    \bx_n+\frac{\bp_n}{m}\int_t^{t+\Delta t}a^{-2}\mathrm{d}t
  \end{cases},\\
K(\Delta t):&~
\begin{cases}
    \displaystyle \bp_n \rightarrow 
    \bp_n+\bff_n\int_t^{t+\Delta t}a^{-1}\mathrm{d}t\\
    \displaystyle \bx_n\rightarrow \bx_n
  \end{cases},
\end{align}
where $\bff_n = -m\nabla_{\bx}\Phi(\bx_n).$
The factors $I_{\rm drift}=\int_t^{t+\Delta t}a^{-2}\mathrm{d}t$ and $I_{\rm kick}=\int_t^{t+\Delta t}a^{-1}\mathrm{d}t$ are called the drift and kick factors, respectively.

\subsection{SU simulation including large-scale tidal effect in an anisotropic expanding background}
\label{sec:ani_sep}

The SU simulation technique is a useful way to include the effect of super-box density fluctuations on cosmic structure formation in an $N$-body simulation \citep{1996ApJS..103....1B,Sirko05,2011ApJS..194...46G,Li14,Wagner15a,2016JCAP...09..007B,schmidt18}. In this method, we can absorb the effect of large-scale gravitational force into the background metric by modifying the scale factor as illustrated in Figure~\ref{fig:coord}. 

We begin with a brief review of the concept behind the SU simulation \citep{TakadaHu13,akitsu17}.
We first consider the gravitational field smoothed by a survey window $W$, $\Psi^L(\bx)$, where the superscript ``$L$'' denotes that the gravitational field arises from the long-wavelength density fluctuations with scales comparable with or greater than a size of the survey window. Here we also introduce the peculiar gravitational potential $\Psi$ obeying the Poisson equation, $\nabla_{\bx}^2 \Psi=4\pi G\bar{\rho}a^2\delta(\bx,t)$, and note that this is different from the definition of the peculiar gravitational potential defined by Eq.~(\ref{eq:peculiar_Poisson_iso}) via 
$\Phi=a\Psi$.
Taylor-expanding the gravitational field around the position of $\bx_0$ (e.g. the center of the survey volume), we have
\begin{align}
\Psi^L(\bx)
&=\Psi^L(\bx_0)
+\nabla_i\Psi^L|_{\bx_0}\Delta x^i
+\frac{1}{2}\nabla_i\nabla_j\Psi^L|_{\bx_0}\Delta x^i \Delta x^j\nonumber\\
&+\mathcal{O}(\nabla^3\Psi^L|_{\bx_0}\Delta x^3)
\label{eq:Psi_expansion}
\end{align}
where $\Delta x^i=(\bx-\bx_0)^i,~\nabla_i=\partial/\partial x_i$. 
The superscript ``$L$'' denotes the gravitational potential that arises from the super-survey modes. 
Without loss of generality we can 
decompose the second derivative tensor into two independent modes, the trace part and the trace-less tensor part as 
\begin{align}
\left.\nabla_i\nabla_j\Psi^L\right|_{\bx_0}
&=4\pi G\bar{\rho} a^2 \left(
\frac{1}{3}\delta_{ij}^{\rm K}\delta_{\rm b}+K_{ij}
\right),
\end{align}
where 
\begin{align}
\delta_{\rm b}&\equiv\frac{1}{4\pi G\bar{\rho}a^2}\left.\nabla^2\Psi^L\right|_{\bx_0}\\
K_{ij}&\equiv \frac{1}{4\pi G\bar{\rho}a^2}\left.\left(\nabla_i\nabla_j\Psi^L-\frac{1}{3}\delta_{ij}^{\rm K}\nabla^2\Psi^L\right)\right|_{\bx_0},
\label{eq:tidal-tensor}
\end{align}
$\delta^{\rm K}_{ij}$ is the Kronecker delta function, $\delta_{\rm b}$ is the density contrast (DC mode)
 in the survey window and $K_{ij}$ is the super-survey tidal tensor satisfying the trace-less condition, 
 ${\rm Tr}(K_{ij})=0$. As can be found from the above equation, $K_{ij}(t)\propto D_+(t)$, where $D_+(t)$ is the linear growth 
factor, if a survey volume is sufficiently large and in this case the gravitational field $\Psi^L$ is safely considered to be in the linear regime.
Since $\delta_{\rm b}$ and $K_{ij}$ are independent in the linear regime, we consider $\delta_{\rm b}=0$ throughout this paper.

For a particular realization of a sufficiently large-volume galaxy survey, the large-scale tidal tensor is only a temporal function, $K_{ij}=K_{ij}(t)$, 
evolving according to the linear growth rate, and the amplitude is randomly drawn from the Gaussian statistics of survey scale 
in the adiabatic Gaussian initial conditions in the standard $\Lambda$CDM model. 
As discussed in \citet{akitsu17}, the effect of $K_{ij}$ can be absorbed by modifying the background expansion of 
the local realization. However, to do this, we need to consider an anisotropic expansion for the local background, because the effect is anisotropic by definition, and cannot be absorbed by modifying an isotropic FRW background, unlike the effect of large-scale density contrast, $\delta_{\rm b}$ 
\citep{Li14}.
Without loss of generality we can take the coordinate system whose axes are along
the principal axes of the large-scale tidal tensor, $K_{ij}$. 
In this coordinate system, 
the tensor $K_{ij}$ becomes diagonal: $K_{ij}=K_i\delta_{ij}^{\rm K}$, where $K_i$ is the eigenvalues along each axis and 
the traceless condition reads $\sum_i K_i=0$. 
Using the Zel’dovich approximation \citep{zeldovich70} \citep[also see][]{1996ApJS..103....1B,akitsu17,akitsu18}, we 
introduce
the anisotropic scale factor
along each coordinate axis, denoted as $a_{Wi}(t)$:
%
\begin{align}
a_{Wi}(t)
\simeq a(t)\left[1-K_i(t)\right]
\equiv a(t)\alpha_{Wi}(t) 
\label{eq:ani_scale}
\end{align}
where $a(t)$ is the scale factor of the global background and obeys the original Friedmann equation (Eq.~\ref{eq:Friedmann}). 
Hereafter quantities with subscript ``$W$'' denote their quantities in the local background 
\smrv{or the comoving coordinate of SU simulation as defined below.}
$\alpha_{Wi}(t)$ is the normalized anisotropic scale factor, defined as $\alpha_{Wi}(t)\equiv a_{Wi}(t)/a(t)$, satisfying the 
condition, 
$\sum_i \alpha_{Wi}=3$. 
For a sufficiently high redshift, $a_{Wi}(t)\rightarrow a(t)$ due to $K_i(t)\rightarrow 0$.
We use the code to calculate $\alpha_{Wi}$ for a general case following the method in 
\citet{schmidt18} (see Section~2 in their paper).

For a sufficiently large volume, the large-scale tidal force is safely in the linear regime. We introduce three 
constants, 
$\lambda_i$ $(i=1,2,3)$, to parameterize the normalization of
large-scale tidal tensor as
%
\begin{align}
\mtrv{
K_i(t)=D_+(t)\lambda_i .}
\end{align}
The traceless nature of $K_{ij}$ leads to $\sum_i \lambda_i=0$. $\lambda_i$ can be considered as normalization parameters of the large-scale 
tidal tensor today as we adopt the normalization  given by $D_{+}(t_0)=1$ today. 

Whilst the physical coordinate should be invariant,
the comoving coordinate in the local background needs to be modified, if the local scale factor $a_W$ is introduced: 
%
\begin{align}
r_i = a_{Wi}(t)x_{Wi},
\label{eq:comoving_coordinate}
\end{align}
where $x_{Wi}$ is the comoving coordinate component in the local coordinate to which we will simply refer as ``local comoving coordinate''. 
The local comoving coordinate
is different from the comoving coordinate in the global background, simply ``global comoving coordinate''. 
For a sufficiently high redshift, $r_i=a_{Wi}(t) x_{Wi} \rightarrow a(t) x_{Wi} \rightarrow a(t)x_i$. That is, the local comoving coordinate 
coincides with the global comoving coordinate ($x_i$);
$x_{Wi}=x_i$ at a sufficiently high redshift, 
which gives the Lagrangian coordinate condition. Accordingly Eq.~(\ref{eq:comoving_coordinate}) means that the comoving wavenumber in the local coordinate is modified as
\begin{align}
k_{Wi}=\frac{a_{Wij}}{a}k_{j}\simeq \left(\delta_{ij}^{\rm K}-K_{ij}\right)k_j=\left(1-K_{i}\right)k_i=\alpha_{Wi} k_i,
\end{align}
where we used the diagonal condition of $K_{ij}$, $K_{ij}=K_i\delta^{\rm K}_{ij}$. Note that the notation 
$\alpha_{Wi}k_{i}$ does not take the summation over $i$. In the following, we explicitly use the notation $\sum_i$ to mean
the summation.
Throughout this paper, we employ the ``growth-dilation'' technique 
to run the 
SU simulation in the local background with anisotropic expansion 
\citep[see Section~IIIC in][for the isotropic SU simulation case]{Li14}. In this technique, the SU simulation 
allows us to numerically compute the effect of the super-survey modes on the ``growth'' of sub-box modes even in a finite-volume simulation. Then the effect of a modification in the comoving coordinate, 
$x_{Wi}$, compared to $x_i$, can be taken into account separately (see below for details).

In the local background, \tnrv{it is convenient to introduce a normalized velocity corresponding to} the canonical momentum 
as
%
\begin{align}
u_{Wi} = a_{Wi}^2\dot{x}_{Wi}.
\label{eq:velocity_u_ani}
\end{align}
In addition to the Newtonian potential $\phi$
sourced by the mass density field $\rho(\bx; t)$~ \tnrv{(Eq.~\ref{eq:force_iso})}, 
we account for the effect of large-scale tidal force by introducing the external gravitational potential, denoted as $\phi_{W{\rm ext}}(\br; t)$:
%
\begin{align}
\phi_{W\mathrm{ext}}(\br; t) \equiv 2\pi G\bar{\rho}(t) D_+(t) \sum_i \lambda_i r_i^2.
\label{eq:ani_exp_pot}
\end{align}
The gravitational force arising from this potential is given as
%
\begin{align}
-\partial_{r_i}\phi_{W\mathrm{ext}} = - 4\pi G\bar{\rho}(t)D_+(t)\lambda_i r_i.
\end{align}
Furthermore, the traceless condition, $\sum_i\lambda_i=0$, reads
%
\begin{align}
\bnr^2\phi_{W\mathrm{ext}} = 0.
\end{align}
Thus, denoting the sum of the two potentials as $\phi_\mathrm{tot} = \phi + \phi_{W\mathrm{ext}}$, the total potential still obeys the same Poisson equation:
%
\begin{align}
\bnr^2 \phi_\mathrm{tot}(\br) = 4\pi G 
\smrv{\rho(\br)}
-\Lambda.\label{eq:Poisson_ani}
\end{align}
In the presence of \mtrv{the large-scale tide}, the mean mass density is given by
\begin{align}
\bar{\rho}_W\!(t)
&\equiv \frac{\bar{\rho}_{0}}{[a_{W1}a_{W2}a_{W3}](t)}\nonumber\\
&\simeq 
\frac{\bar{\rho}_{0}}{a(t)^3\left[1-D_+\sum_i\lambda_i+O((D_+\lambda_i)^2)\right]}\nonumber\\
&=\frac{\bar{\rho}_{0}}{a(t)^3\left[1+O((D_+\lambda_i)^2)\right]}.
\label{eq:density_ani}
\end{align}
For a sufficiently high redshift, 
$\smrv{\bar{\rho}_W\!(t)}\simeq
\bar{\rho}_{0}/a(t)^3$ because $D(t)\lambda_i \rightarrow 0$
and there $a_{Wi}(t)\simeq a(t)$. Note that $\bar{\rho}_W\!(t)\ne \bar{\rho}(t)$ at the order $O((D_+\lambda)^2)$.

Consider a test particle that rests in the local comoving coordinate.
As $\dot{x}_{Wi}=0$ for such a particle, the acceleration of the particle is computed as
%
\begin{align}
\ddot{r}_i = \ddot{a}_{Wi}x_{Wi}.
\label{eq:r_awxw}
\end{align}
The gravitational
force on this particle can be computed as a sum of the gravitational force arising from the spatially-homogeneous parts, i.e. the mean mass density ($\bar{\rho}$), the cosmological constant ($\Lambda$)
and the large-scale tidal tensor ($K_{ij}$ or $D_+\lambda_i$): 
%
\begin{align}
\ddot{r}_i &= -\bnr\bar{\phi} - \bnr\phi_{W\mathrm{ext}},\nonumber\\
&=\left[-\frac{4\pi G\bar{\rho}_W(t)}{3}+\frac{\Lambda}{3}-4\pi G\bar{\rho}(t)D_+(t)\lambda_i\right]r_i.
\end{align}
Thus, very similarly to the isotropic expanding background in the $\Lambda$CDM model, we can account for this 
homogeneous gravitational force by introducing the scale factor in the local background obeying the following time-differential equation (Figure~\ref{fig:coord}): 
%
\begin{align}
\frac{\ddot{r}_i}{r_i}=\frac{\ddot{a}_{Wi}}{a_{Wi}} =  -\frac{4\pi G\bar{\rho}_W(t)}{3}+\frac{\Lambda}{3}
-4\pi G\bar{\rho}(t)D_+(t)\lambda_i ~.\label{eq:ddot_a_ani}
\end{align}
This is an effective Friedmann equation \mtrv{to describe an anisotropic expansion due to the large-scale tidal effect
in the local background}, and consistent with Eq.~(14) in \cite{schmidt18} in the linear regime ($D_+\lambda_i\ll 1$).
It is useful to notice that, after summing 
the three components of the above equation, i.e. $\sum_i$, it recovers the 
isotropic Friedmann equation due to
$\sum_i\lambda_i=0$ in \smrv{the}
angle-average sense \smrv{or for the monopole component of $a_{Wi}$.}

Following the similar concept for the isotropic background case discussed around Eqs.~(\ref{eq:peculiar_acceleration})
and (\ref{eq:peculiar_Poisson_iso}), we introduce the peculiar acceleration along each coordinate axis in the local background as
%
\begin{align}
\dot{u}_{Wi} = -\partial_{x_{Wi}}\phi_\mathrm{tot} - a_{Wi}\ddot{a}_{Wi} x_{Wi}.
\end{align}
A sensible choice for the peculiar potential to absorb the second term is
%
\begin{align}
\Phi_W=a\left[\phi_{\rm tot}+\frac{1}{2}\sum_i a_{Wi} \ddot a_{Wi} x_{Wi}^2\right],\label{eq:new_pot}
\end{align}
This is consistent with the Hamiltonian (see Sec.~\ref{sec:Hami}) as
%
\begin{align}
\dot u_{Wi}=-\frac{1}{a}\frac{\partial \Phi_{W}}{\partial x_{Wi}}.\label{eq:force_ani}
\end{align}
Since the peculiar gravitational potential defined above arises form the {\it inhomogeneous} matter distribution, 
$\nabla_{\bx}^2\Phi_W=4\pi G\bar{\rho}_0\delta(\bx,t)$.
Our tasks are to express the peculiar gravitational potential in terms of variables in the local background with anisotropic expansion, $a_{Wi}$. From Eq.~(\ref{eq:r_awxw}), the Poisson equation for the peculiar gravitational potential is expressed as
%
\begin{align}
\sum_i\left(\frac{1}{\alpha_{Wi}}\frac{\partial}{\partial x_{Wi}}\right)^2\Phi_W
= \frac{4\pi G\bar{\rho}_{0} \delta}{\alpha_{W1}\alpha_{W2}\alpha_{W3}}. 
\label{eq:Poisson_ani_new}
\end{align}
We rewrite this in Fourier space for the PM force calculation as
%
\begin{align}
\Tilde{\Phi}_{W,\bk_W}
&=-\frac{4\pi G \bar{\rho}_{0} \Tilde{\delta}_{\bk_W}}{\alpha_{W1}\alpha_{W2}\alpha_{W3}\sum_i (k_{Wi}/\alpha_{Wi})^2}\nonumber\\
&=-\frac{4\pi G \bar{\rho}_0 \Tilde{\delta}_{\bk_W}}{\alpha_{W1}\alpha_{W2}\alpha_{W3}}G_W(k_W),
\end{align}
where $\sum_i (k_{Wi}/\alpha_{Wi})^2 = \sum_i k_i^2=k^2$ is the square of global isotropic comoving wavenumber reexpressed in terms of that in the anisotropic ``simulation'' coordinates, and $G_W(k_W)\equiv 1/[\sum_i (k_{Wi}/\alpha_i)^2]$ is the 
Green function, which is also kept unchanged when it is expressed by the wavenumber
in the global comoving coordinate, 
$G_W(k_W)\rightarrow G(k)=1/k^2$.
This suggests that we need to modify the Green function in the local comoving coordinate, i.e. the SU simulation coordinate. 
We do not have to change the density assignment for the PM force calculation.
Thus we keep the grid for the density assignment to be equally spaced in the local comoving coordinate.
We also need to express
the split function between the PM and the Tree forces in terms of the local comoving coordinate variables.
We use the function which splits isotropically in the global comoving coordinate:
\begin{align}
\exp\left[-\sum_i(k_{Wi}/\alpha_{Wi})^2~ \xs^2\right].
\end{align}
Again, because of $\sum_i(k_{Wi}/\alpha_{Wi})^2=k^2$, 
this recovers the split function in the global comoving coordinate. $\xs$ is a constant, and we set $x_s=4.5 L_{\rm box}/N_{\rm grid}^{1/3}$ in the SU simulation. 

\subsection{Tree force in an anisotropic SU simulation}

In the anisotropic case, the Tree acceleration is
\begin{align}
&\left.-\frac{\partial \Phi_W^{\rm T}}{\partial x_{Wi}}\right|_{\bx_{Wn}}
=
-Gm\alpha_{Wi}\sum_{n';n'\neq n} 
\frac{x_{n,i}-x_{n',i}}{|\bx_n-\bx_{n'}|^3}\nonumber\\
&\hspace{4em}\times\left[
\erfc\left(\frac{|\bx_n-\bx_{n'}|}{2\xs}\right)
+\frac{|\bx_n-\bx_{n'}|}{\xs\sqrt{\pi}}\exp\left(-\frac{|\bx_n-\bx_{n'}|^2}{4\xs^2}\right)
\right].
\label{eq:tree_ani_resc2}
\end{align}
Note that variables on the r.h.s. 
must be rewritten in terms of variables in the local comoving coordinate \tnrv{in the actual code implementation}, e.g. 
$x_{n,i}=\alpha_{Wi} x_{Wn,i}$ and $|\bx_n-\bx_{n'}|=\left[\sum_i\left\{(\alpha_{Wi})^2(x_{Wn,i}-x_{Wn',i})^2\right\}\right]^{1/2}$.
In the case of Eq.~(\ref{eq:tree_ani_resc2}), the Tree force is cut isotropically in the 
global comoving
coordinate consistently to
the PM force.
Comparing to Eq.~(\ref{eq:tree_iso_com}), we need to 
multiply an additional factor of $\alpha_{Wi}$ to take account the difference between $\bx$ and $\bx_W$ in the derivative.
We use the softening kernel form same as the standard isotropic case.
Thus the final form of the Tree acceleration in the anisotropic case is
\begin{align}
\left.-\frac{\partial \Phi^{\rm T}}{\partial x_{Wi}}\right|_{\bx_{Wn}}
&=
G\alpha_{Wi}\sum_{\rm group} M_{\rm group}g_1(y_{\rm group})y_{{\rm group},i}\nonumber\\
&\times\left[
\erfc\left(\frac{y_{\rm group}}{2\xs}\right)
+\frac{y_{\rm group}}{\xs\sqrt{\pi}}\exp\left(-\frac{y_{\rm group}^2}{4\xs^2}\right)
\right],
\end{align}
where $y_{\rm group}=\left[\sum_{i}\left\{(\alpha_{Wi})^2(x_{Wn,i}-s_{W{\rm group},i})^2\right\}\right]^{1/2}$ and so on.

\subsection{The drift and kick factors from the anisotropic Hamiltonian}
\label{sec:Hami}

We consider the Hamiltonian for a system of $N$-body particles
in the local background with anisotropic expansion. 
The physical velocity of the $n$-th particle in the local background is given as
%
\begin{align}
\frac{\mathrm{d}(a_{Wi} x_{Wn,i})}{\mathrm{d}t}
\simeq\alpha_{Wi} \left(a\frac{\mathrm{d}x_{Wn,i}}{\mathrm{d}t}+x_{Wn,i}\frac{\mathrm{d}a}{\mathrm{d}t}\right),
\end{align}
where we have assumed that $\alpha_i$ is nearly constant during an infinitesimal time interval
\citep[see][]{schmidt18}.
The first term is the peculiar velocity and the second one is the Hubble flow in the local coordinate.
The canonical momentum is 
\begin{align}
p_{Wn,i}=m a_{Wi}^2\dot x_{Wn,i}.
\end{align}
The kinetic term of the Hamiltonian $H_{W\rm kin}$ is
\begin{align}
H_{W\rm kin}=\sum_n\sum_i\frac{p^2_{Wn,i}}{2 m a^2_{Wi}}.
\end{align}
The potential term is
\begin{align}
H_{W\rm pot}
=\frac{1}{2}\sum_{n}\frac{m\Phi_W(\bx_{Wn})}{a}.
\end{align}
The total Hamiltonian in the local background that describes the gravitational system for the spatially-inhomogeneous matter distribution
is
\begin{align}
H_W=\sum_n\sum_i\frac{p^2_{Wn,i}}{2 m a^2_{Wi}}
+\frac{1}{2}\sum_{n}\frac{m\Phi_W(\bx_{Wn})}{a}.
\end{align}

The resulting canonical equations of motion are
\begin{align}
\dot x_{Wn,i}
&=\frac{\partial H_W}{\partial p_{Wn,i}}
=\frac{1}{a_{Wi}^2}\frac{p_{Wn,i}}{m},\\
\dot p_{Wn,i}
&=-\frac{\partial H_W}{\partial x_{Wn,i}}
=-\frac{m}{a}\frac{\partial \Phi_W(\bx_{Wn})}{\partial x_{Wn,i}}
\end{align}
As we do in the isotropic case, we obtain the drift and kick factors employed as the time-evolution operators in {\sc Gadget-2} for 
the SU simulation case:
\begin{align}
I_{{\rm drift},~i}&=\int_t^{t+\Delta t}\frac{\mathrm{d}t}{a_{Wi}^2}\simeq\frac{1}{\alpha^2_{Wi}}\int_t^{t+\Delta t}\frac{\mathrm{d}t}{a^2},\\
I_{\rm kick}&=\int_t^{t+\Delta t}\frac{\mathrm{d}t}{a},
\end{align}
i.e., the drift factor becomes anisotropic and the kick factor remains isotropic in the local coordinate (SU simulation coordinate).
We assumed $\alpha_{Wi}$ to be constant during a time step adopted in the simulation.

\subsection{Summary: SU simulation implementation}

\mtrv{In summary we made the following changes in the {\sc Gadget-2} code to implement the SU simulation including the effect of super-box tidal force into the local background with anisotropic expansion.}
\begin{itemize}
	\item solve the effective Friedmann equation (Eq.~\ref{eq:ddot_a_ani}) to obtain the scale factor in the local background, $a_{Wi}(t)$, along each coordinate axis that describes an anisotropic expansion. 
	In this paper we use the publicly available code given in \citet{schmidt18} to obtain $a_{Wi}(t)$ that is accurate up to $O(K^2)$. 
	If $|K_i|=D_+|\lambda_i|\ll 1$ holds during epochs of a simulation, we can find that
 $a_{Wi}(t)=a(t)\alpha_{Wi}(t) \simeq a(t) (1-D_+\lambda_i)$ is a solution, for the growth rate $D_+(t)$ in the global background cosmology and 
 for the input tidal tensor, $\lambda_i$. 
     \item the PM part of the peculiar gravitational force that is sourced by the matter density fluctuation field 
     (spatially-inhomogeneous source): 
    \begin{itemize}
        \item the Green function: $1/[\sum_i k^2_i]\rightarrow 1/[\sum_i (k_{Wi}/\alpha_{Wi})^2]$
        \item the cosmic mean density: $\bar\rho_0\rightarrow \bar\rho_0/(\alpha_1\alpha_2\alpha_3)$
        \item the split factor: $\exp\left(-\bk^2~\xs^2\right)\rightarrow
        \exp\left[-\sum_i (k_{Wi}/\alpha_{Wi})^2~ \xs^2\right]$
    \end{itemize}
    \item the Tree part:
    \SM{
    \begin{itemize}
        \item multiply an additional factor \smrv{$\alpha_{Wi}$} to the acceleration
        \item the positions and the distances: e.g., $x_{n,i}\rightarrow\alpha_{Wi} x_{Wn,i}$, $|\bx_n-\bx_{n'}|\rightarrow\left[\sum_i\left\{(\alpha_{Wi})^2(x_{Wn,i}-x_{Wn',i})^2\right\}\right]^{1/2}$
    \end{itemize}
    }
    \item the drift factor: $\int_t^{t+\Delta t}a^{-2}\mathrm{d}t\rightarrow
    \alpha^{-2}_{Wi}\int_t^{t+\Delta t}a^{-2}\mathrm{d}t$
\end{itemize}
Our implementation of SU simulations differs from that in \citet{schmidt18} for details of the numerical methods, and therefore our method gives another consistency test of SU simulations.
In particular, the kick factor in \citet{schmidt18} is anisotropic as $\alpha_{Wi}^{-1}\int_t^{t+\Delta t}a^{-1}\mathrm{d}t$.

\section{The response function of matter power spectrum to the large-scale tidal force}
\label{sec:GK}

In this paper we use the SU simulations including the large-scale tidal effect to calibrate the ``response'' function of the
matter power spectrum, which describes how the matter power spectrum of sub-box modes responds to the large-scale tidal force
\citep{2015JCAP...10..059D,2017JCAP...02..025I,akitsu17,schmidt18,Akitsu19}.  

\subsection{Growth and dilation responses}

For a finite volume survey, we can sample only sub-survey modes of the fluctuations, and can measure 
the statistics, here mainly focused on the matter power spectrum, measured from the local volume. The super-survey tidal force, which is not a direct observable, affects
the matter power spectrum in the local volume 
through the mode coupling 
in nonlinear structure formation. 
As a result, the band power measured from the local volume is modulated 
depending on the degree of alignments between 
$\bk$ and the large-scale tidal tensor \citep{akitsu17}. That is, the power spectrum 
acquires a dependence
on the direction of $\bk$, 
in addition to the
length $|\bk|$: $P(\bk;  K)$ \smrv{in the presence of the super-survey tidal effect, $K_{ij}$.}
Using the Taylor expansion we can express $P(\bk; K)$
to the first-order of $K_{ij}$ as
\begin{align}
P(\bk;K)
&\simeq P(\bk;K=0)
+\left.\frac{d P(\bk;K)}{d K_{ij}}\right|_{K=0} K_{ij}\nonumber\\
&=P(k)\left[
1+R_K(k)\hat k_i \hat k_j K_{ij}
\right],
\label{eq:def-res}
\end{align}
where $P(\bk;K_{ij}=0)$ is the power spectrum for the global background in the absence of the super-survey tidal effect, which therefore depends only on the length of wavenumber, 
i.e. $P(\bk;K_{ij}=0)=P(k)$. \mtrv{Here we introduced
the so-called {\it response function, $R_K(k;t)$, } 
that describes a response of the matter power spectrum to the super-survey tidal tensor, defined as}
\begin{align}
    R_K(k; t)\hat k_i \hat k_j
    \equiv \frac{1}{P(k)}\left.\frac{\mathrm{d} P(\bk;K)}{\mathrm{d} K_{ij}}\right|_{K=0},
    \label{eq:def_RK}
\end{align}
and $\hat{\bk}$ is the normalized $k$ vector as $\hat{k}_i=k_i/\sqrt{\sum_i k_i^2}$. The second term in Eq.~(\ref{eq:def-res}) describes a modulation in the power spectrum measured from a local volume 
under the super-survey tidal tensor $K_{ij}$.

The SU simulation is equivalent to the case that structure formation is simulated in the global background with a sufficiently large volume, including long-wavelength perturbations, and then the power spectrum is measured from a sub-volume of the large-scale simulation box, corresponding to the volume of SU simulation. Since we do not consider the super-survey density contrast, i.e. $\delta_{\rm b}=0$, we have $\bar{\rho}_W=\bar{\rho}$ at the first order of $K_{ij}$ and the Hubble constant in the local volume is not modified, $h_W=h$, unlike the case for $\delta_{\rm b}$ in \citet{Li14}. 
Hence, even if the power spectrum has a dimension of [$(h^{-1}{\rm Mpc})^3$], the relation $P(\bk, K)=P_W(\bk_W, K)$ holds, where 
$k_{Wi}=k_j(\delta_{ij}^{\rm K}-K_{ij})$ and $P_W$ is the same power spectrum in the local volume but it takes the wavevector in the local comoving coordinate in the first argument. 
Following the method in \citet{Li14} (see around Eq.~47 in their paper), we can use the chain rule to compute the derivative in Eq.~(\ref{eq:def_RK}) to the first order of $K_{ij}$ as 
\begin{align}
&\left.
\frac{\mathrm{d}  P(\bk, K)}{\mathrm{d} K_{ij}}
\right|_{\bk,{K=0}}\nonumber\\
&=\left.\frac{\partial  P_W(\bk_W,K)}{\partial K_{ij}}\right|_{\bk_W,K=0}+
\frac{\partial  P_W(\bk_W,K=0)}{\partial k_{Wm}}
\frac{\partial k_{Wm}}{\partial K_{ij}}\nonumber\\
&\simeq
\left.\frac{\partial  P_W(\bk_W,K)}{\partial  K_{ij}}\right|_{\bk_W,{K=0}}+
\frac{\partial  P(k)}{\partial  k} \frac{\partial k}{\partial k_{m}}\frac{\partial k_{Wm}}{\partial K_{ij}}\nonumber\\
&=
\left.\frac{\partial  P_W(\bk_W,K)}{\partial  K_{ij}}\right|_{\bk_W,K=0}-
\frac{\partial  P(k)}{\partial  \ln k} \hat{k}_i\hat{k}_j.
\end{align}
The first term on the r.h.s. of the above equation is the ``growth'' response that describes how the growth of the density perturbation, $\delta_{\bk}$, is affected by the large-scale tidal force.  The second term denotes the ``dilation'' response; 
the modulation in the power spectrum arises from the fact that the comoving wavenumber in the local volume 
is modulated by the large-scale tidal force via $k_{Wi}=k_j(\delta_{ij}^{\rm K}-K_{ij})$.
Thus we define the growth response as
\begin{align}
G_K(k)\hat{k}_i\hat{k}_j\equiv \frac{1}{P(k)}\left.\frac{\partial P_W({\bk_W},K)}{\partial K_{ij}}\right|_{\bk_W,{K=0}} ~ .
\end{align}
Then $R_K$ and $G_K$ are related as
\begin{align}
    \SM{R_K(k)=G_K(k)-\frac{\partial \ln P(k)}{\partial \ln k}.}
    \label{eq:RK-GK}
\end{align}
Thus the SU simulation gives a useful way to calibrate the growth response. We can consider the dilation response separately from the derivative of the nonlinear power spectrum. 
The perturbation theory  prediction $G_K=8/7$ \citep{akitsu17,BarreiraSchmidt17} can be used to test our anisotropic simulation code at small $k$ in the linear regime.

To compute the growth response, we will measure the following power spectrum measured in the SU simulation \citep[see around Eq.~25 in][]{Takahashi19}: 
\begin{align}
\hat{P}_W(k_{W,{\rm bin}},K)=\frac{1}{N(k_{W,{\rm bin}})}\sum_{\bk_W\in k_{W,{\rm bin}}}|\delta_{\bk_W}|^2. 
\label{eq:est_Pkw}
\end{align}
Here the power spectrum is estimated from the average of $|\delta_{\bk_W}|^2$ over a bin given by a 
spherical shell 
at 
radius around $k_{W,{\rm bin}}$ 
with a finite width. Recalling the relation between the wavevectors in the global and local backgrounds, 
$k_{Wi}=k_i(1-D_+\lambda_i)=\alpha_{Wi}k_i$, the power spectrum estimated in the above is equivalent to the power spectrum estimated from the average of 
$|\delta_{\bk}|^2$
over the ellipsoidal shell of $\bk$ satisfying 
$k_{W
}^2=\sum_{i}(\alpha_{Wi})^2k_i^2$ 
when it is seen in the global comoving coordinate. 
Hence the above estimator gives an estimate of $\bk$-direction dependent power spectrum which we want for the response calibration: $P(\bk, K)$.  
$N(k_W)$ is the number of Fourier modes taken in the summation, which is given by $N(k_{W,{\rm bin}})\simeq 2\pi k_{W,{\rm bin}}^2\Delta k_W V/(2\pi)^3$ at $k_{W,{\rm bin}}\gg 2\pi/L$, where $\Delta k_W $ is the bin width and $V $ is the volume of SU simulation box.

\subsection{A calibration method of the growth response}

We follow \citet{schmidt18} to measure the growth response 
$G_K$ from the power spectra measured from the SU simulations (Eq.~\ref{eq:est_Pkw}).
To do so, we employ the three simulations labeled A, B and C.
They share
the same 
random-number seed for the generation of the initial conditions, but different $\lambda$ values: $\blambda_{\rm A}=\lambda_{{\rm A},z}(-0.5,-0.5,1),~\blambda_B=\lambda_{{\rm B},z}(-0.5,-0.5,1)=\lambda_{{\rm A},z}(0.5,0.5,-1)=-\blambda_{\rm A}$ and $\blambda_{\rm C}=(0,0,0)$.
Thus the A and B runs are the anisotropic case and $\delta_{\rm b}=0$ due to $\sum_i \lambda_i=0$, and the C run is the standard isotropic case.

The estimator of $G_K$ is
\begin{align}
G_K(k)=\frac{\langle[P_{W\rm A}(\bk_{W\rm A})
\tnrv{\mathcal{L}_2(\hat k_{W \mathrm{A},z})}
-P_{W\rm B}(\bk_{W\rm B})\tnrv{\mathcal{L}_2(\hat k_{W \mathrm{B},z})} 
]
\rangle}
{\langle P_{W\rm C}(\bk_{W\rm C})\mathcal{L}_2^2(\hat k_z)D_+(t)(\lambda_{{\rm A},z}-\lambda_{{\rm B},z}) \rangle},
\label{eq:est_GK}
\end{align}
where $\langle ...\rangle$ is angle averaging, $\mathcal{L}_2$ is the second-order Legendre polynomial.
$P_{W\rm A}(\bk_{W\rm A}),~P_{W\rm B}(\bk_{W\rm B})$ and $P_{W\rm C}(\bk_{W\rm C})$ are the three dimensional matter power spectrum  measured in each simulation frame for the runs of A, B and C, respectively.
Note that each simulation and measured $G_K$ can be characterized by a single parameter $\lambda_z$ and $\lambda_{{\rm A},z}$, respectively.
Throughout the paper, we frequently use these parameters to present results.

\section{SU simulations}
\label{sec:sim}

Here we describe a numerical implementation of 
SU simulations. We first 
describe
how to generate the initial conditions, and then present the specifications of our simulations.
We also study the resolution effects on measurements of the power spectrum from SU simulations. 

\subsection{Generating initial conditions}

To generate the initial conditions, we use the second-order Lagrangian perturbation theory \citep[2LPT;~][]{scoccimarro98,crocce06a,nishimichi09} and the linear matter power spectrum computed using
{\sc CAMB}  \citep{camb}.
Recalling the fact that 
the growth tidal response function $G_K$ is $8/7$ at a sufficiently high redshift or 
at \mtrv{the leading order of the perturbation  theory prediction} \citep{akitsu17,BarreiraSchmidt17}, we change the input linear matter power spectrum, for an assumed large-scale tidal tensor $K_{ij}=D_+(a_{\rm ini})\lambda_i$ at the initial redshift, as
\begin{align}
P(\bk, a_{\rm ini})
\rightarrow P_W(\bk_W, a_{\rm ini})=P(k, a_{\rm ini})\left[1+\frac{8}{7}\hat{k}_i\hat{k}_jK_{ij}(a_{\rm ini})\right].
\label{eq:pk-ic}
\end{align}
We use this modified matter power spectrum to calculate 
the initial displacement of particles.

To set up the initial conditions, we also need the initial peculiar velocity field. Rather than analytically deriving the peculiar velocity field in the SU simulation setups, we numerically evaluate it as follows. Recall that the peculiar velocity field 
for an $N$-body particle in the local coordinate is given as
\begin{align}
\frac{\mathrm{d}(a_{Wi}x_{Wi})}{\mathrm{d}t}
\simeq \alpha_{Wi}\frac{\mathrm{d}(ax_{Wi})}{\mathrm{d}t}
= a_{Wi}H\left(x_{Wi}+\frac{\mathrm{d}x_{Wi}}{\mathrm{d}\ln a}\right),
\end{align}
where we ignored a temporal variation of the normalized scale factor $\alpha_{Wi}(t)$ during the 
time step in a simulation.
Hence following the normalization convention of the input and output data for {\sc Gadget-2}, we calculate the peculiar velocity 
\begin{align}
\sqrt{a_{Wi}}H\frac{\mathrm{d}x_{Wi}}{\mathrm{d}\ln a}.
\end{align}
We first prepare the initial displacement fields at two epochs, $a_{{\rm ini}+}$ and $a_{{\rm ini}-}$, slightly shifted from the initial redshift on both positive and negative sides, using the same initial seed of the fluctuations from the initial power spectrum (Eq.~\ref{eq:pk-ic}). 
Then we compute the initial peculiar velocity field for each particle from the numerical differentiation as
\begin{align}
\sqrt{a_{Wi}(a_{\rm ini})}H(a_{\rm ini}) \frac{x_{Wi}(a_{{\rm ini} +})-x_{Wi}(a_{{\rm ini}-})}
{\ln a_{{\rm ini}+}-\ln a_{{\rm ini}-}}.
\end{align}
In this paper we adopt $\mathrm{d}\ln a=\ln a_{{\rm ini}+}-\ln a_{{\rm ini}-}=0.02$ for all the simulations. 
We have tested the accuracy of this step size in the finite differencing
by applying it to the isotropic case where the velocities can directly be computed by 2LPT, 
finding that the errors are less than 5\% for more than 97\% of particles in our fiducial high resolution runs (see Sec.~\ref{sec:sim-spec}).

\subsection{SU simulation specifications}
\label{sec:sim-spec}

\begin{table}
\caption{Summary of our simulation specifications, where $N_{\rm part},~z_{\rm ini}$ and $(\lambda_x,~\lambda_y,~\lambda_z)$ 
denote
the number of particles, the initial redshift and the amplitudes of the super-box tidal tensor at present, respectively.}
\label{tab:sims}
\begin{center}
\begin{tabular}{l||lllll}
\hline\hline
Set & $N_{\rm part}$ & $z_{\rm ini}$ & $(\lambda_x,~\lambda_y,~\lambda_z)$ & Realizations
\\ 
\hline
HR1-A & $512^3$ & 127 & $(-0.05, -0.05, 0.1)$ & 16\\
HR1-B & $512^3$ & 127 & $(0.05, 0.05, -0.1)$ & 16\\
HR2-A & $512^3$ & 127 & $(-0.005, -0.005, 0.01)$ & 16\\
HR2-B & $512^3$ & 127 & $(0.005, 0.005, -0.01)$ & 16\\
HR3-A & $512^3$ & 127 & $(-0.0005, -0.0005, 0.001)$ & 16\\
HR3-B & $512^3$ & 127 & $(0.0005, 0.0005, -0.001)$ & 16\\
HR-C & $512^3$ & 127 & $(0, 0, 0)$ & 16\\
MR-A & $256^3$ & 63 & $(-0.005, -0.005, 0.01)$ & 16\\
MR-B & $256^3$ & 63 & $(0.005, 0.005, -0.01)$ & 16\\
MR-C & $256^3$ & 63 & $(0, 0, 0)$ & 16\\
LR-A & $128^3$ & 31 & $(-0.005, -0.005, 0.01)$ & 16\\
LR-B & $128^3$ & 31 & $(0.005, 0.005, -0.01)$ & 16\\
LR-C & $128^3$ & 31 & $(0, 0, 0)$ & 16\\
\hline\hline
\end{tabular}
\end{center}
\end{table}
Table~\ref{tab:sims} summarizes our simulation specifications, where $N_{\rm part}$ is the number of particles, $z_{\rm ini}$ is the initial redshift and $(\lambda_x,~\lambda_y,~\lambda_z)$ are
the amplitudes of the tidal tensor normalized to their values today, 
respectively. \mtrv{For a validation on our choice of the initial redshift $z_{\rm ini}$,}
we would like to refer readers to
\citet{nishimichi19}.
As we stated, we set the values of $\lambda_x,~\lambda_y$ and $\lambda_z$ such that $\lambda_x+\lambda_y+\lambda_z=0$ and $\lambda_x=\lambda_y=-\lambda_z/2$ for the anisotropic cases, and $\lambda_x=\lambda_y=\lambda_z=0$ for the isotropic cases.
Each simulation set has 16 realizations, and 
each pair of SU--A/B simulations  and the corresponding (C) isotropic simulation
share the identical set of 16 random seeds to generate the initial conditions to minimize the effect of sample variance in the response calibration.
For each run, we adopt a flat $\Lambda$CDM cosmology with $\Omega_{\rm m0}=0.3156,~\Omega_\Lambda=0.6844,~H_0=100h=67.27~{\rm km~s^{-1}~Mpc^{-1}},~n_{\rm s}=0.9645$ and $A_{\rm s}=2.2065\times10^{-9}$ 
for the global background
\citep{planck-collaboration:2015fj}.
We carry out all the simulations 
using
the TreePM mode of the modified {\sc Gadget-2} 
for 
the local comoving simulation box size of $L_{\rm box}=125~h^{-1}~{\rm Mpc}$ on a side, which is sufficiently large so that the 
super-survey tidal force
is safely considered to be in the linear regime.
We set the gravitational softening parameter to be $\epsilon=0.05\times L_{\rm box}/N_{\rm part}^{1/3}=12.2,~24.4$ and $48.8~h^{-1}~{\rm kpc}$ for the sets with $N_{\rm part}=512^3$ (HR), $256^3$ (MR) and $128^3$ (LR), respectively.
We employ $N_{\rm grid}$ Fourier grids for 
the PM force calculation, 
and set this parameter to be $2^3\times N_{\rm part}$ 
(i.e., two grids per particle spacing in one dimension).

\subsection{Resolution study}
\label{sec:resolution}
\begin{figure}
\begin{center}
\includegraphics[width=\hsize]{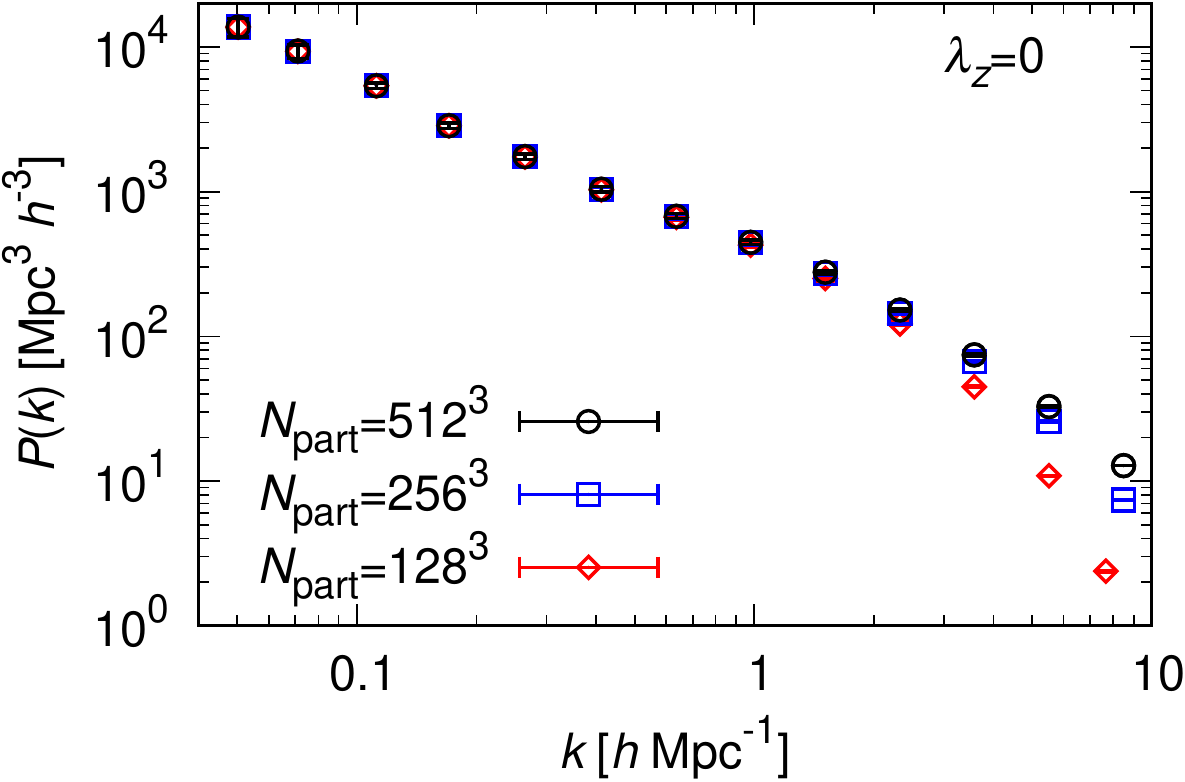}
\caption{The matter power spectra at $z=0$ measured from the simulation sets with 
$N_{\rm part}=512^3$ (HR-C), $256^3$ (MR-C), $128^3$ (LR-C) in the isotropic cases with $\lambda_z=0$. Comparing the three results manifests the validation range of scales up to which the simulation with respective $N_{\rm part}$ is reliable.
The error bars at each $k$ bin denote the error on the mean, estimated from 16 realizations.}
\label{fig:pk-npart}
\end{center}
\end{figure}

In this subsection we address the validation range of scales where simulation results are considered well converged and reliable.
Figure~\ref{fig:pk-npart} shows the matter power spectra at $z=0$ measured 
from 
the sets of HR-C, MR-C and LR-C, i.e., the isotropic cases with $N_{\rm part}=512^3,~256^3$ and $128^3$, 
respectively.
Throughout this paper we show
the error bars to  
denote errors on the mean  that are estimated from
the standard deviation of 16 realizations divided by $\sqrt{16}$.
The figure shows that the simulations of different resolutions start to deviate from each other on scales greater than 
a certain $k$, due to the resolution limitation. 
The LR-C (MR-C) set agrees with the HR-C within 10\% up to $k\simeq1.5~(3)~h~{\rm Mpc}^{-1}$.
Considering 
the size of the softening length, we 
conclude
that the HR sets are \mtrv{likely} reliable 
up to 
$k\simeq 6~h~{\rm Mpc}^{-1}$.

\begin{figure}
\begin{center}
\includegraphics[width=\hsize]{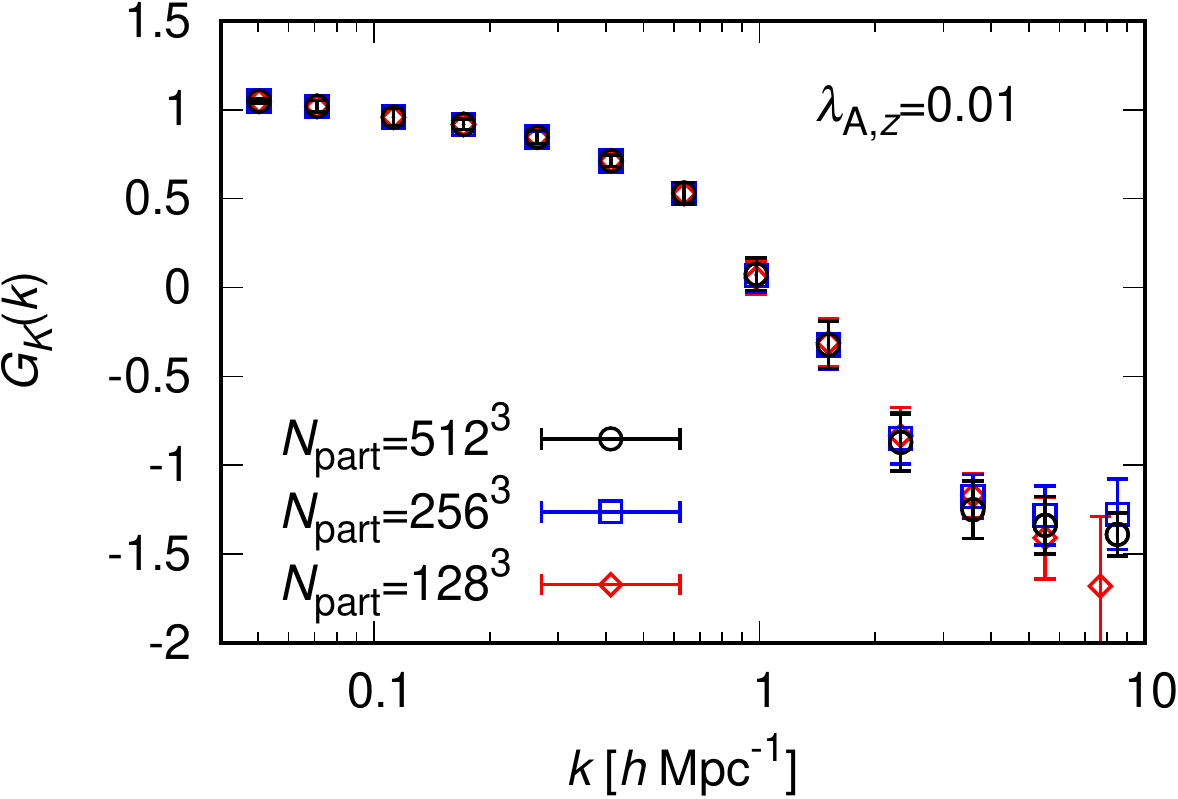}
\caption{The growth response, $G_{K}(k)$, at $z=0$, measured from SU simulations with
$\lambda_{z}= \pm 0.01$ and $N_{\rm part}=512^3$ (HR2), $256^3$ (MR), $128^3$ (LR).
We used the estimator (Eq.~\ref{eq:est_GK}) to evaluate $G_K$ from each paired SU simulations (see Table~\ref{tab:sims}).}
\label{fig:gk-npart}
\end{center}
\end{figure}
Figure~\ref{fig:gk-npart} shows the growth-only tidal response function $G_K$ at $z=0$ for the three resolutions.
For $N_{\rm part}=512^3$, we use a combination of the sets of HR2-A, HR2-B and HR-C to measure $G_K$ using the estimator Eq.(\ref{eq:est_GK}).
Similarly, the case of $N_{\rm part}=256^3~(128^3)$ uses the MR-A, MR-B and MR-C (LR-A, LR-B and LR-C) sets.
The $\lambda$ values are common among the three resolutions as $(\lambda_x,~\lambda_y,~\lambda_z)=(\mp0.005, ~\mp 0.005,~\pm0.01)$ and $(0,~0,~0)$.
This setting can be characterized by a single parameter $\lambda_{{\rm A},z}=0.01$.
The response function
$G_K$ measured from the three resolutions agree with each other within the error bars over 
the whole range of scales we plot here.
This implies that the 
impact of the limited resolution on the three dimensional power spectrum is very similar for the isotropic and anisotropic cases, and this partially cancels in the estimator of the response function.
However, recalling Figure \ref{fig:pk-npart}, we use the HR sets ($N_{\rm part}=512^3$) as our fiducial choice
to study the results up to
$k
\simeq
6~h~{\rm Mpc}^{-1}$ to be on the safe side.

{\color{black}{
\subsection{The impact of the periodic boundary conditions}
\label{sec:su_periodic}
\begin{figure}
\begin{center}
\includegraphics[width=\hsize]{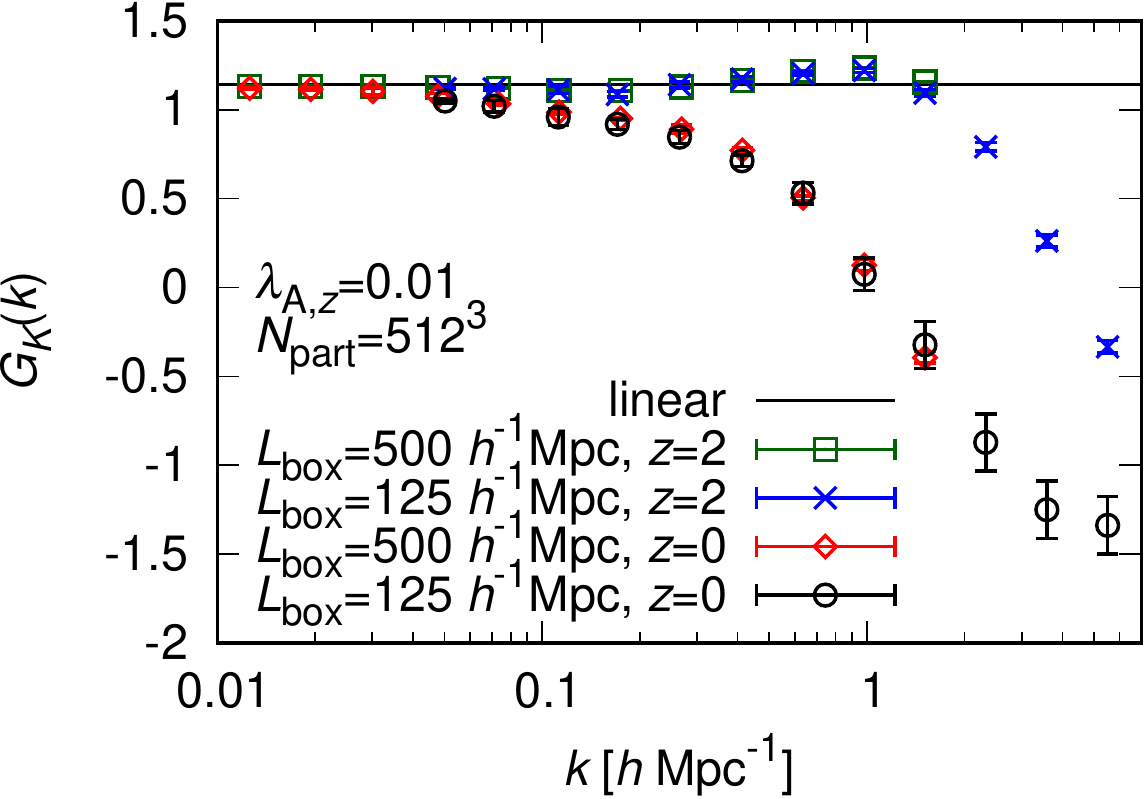}
\caption{The growth response functions $G_K(k)$ at $z=0,2$ from the SU simulations with $L_{\rm box}=125,500~\hiMpc$, $N_{\rm part}=512^3$ and $\lambda_{{\rm A},z}=0.01$.
The solid line shows the perturbation theory prediction of $G_K=8/7$.
}
\label{fig:gk-Lbox}
\end{center}
\end{figure}
In this section we study the impact of the periodic boundary condition, i.e. a possible artifact due to the choice of the simulation box size.
Our anisotropic SU simulations incorporate the effect of the
super-box tidal force
with wavelengths {\it much longer} than a box size $L_{\rm box}$. This is equivalent to an approximation to ignore all the higher-order terms in the long-wavelength gravitational potential; that is, we ignore the contributions of $O(\nabla^n\Psi^L|_{{\bf x}_0}(\Delta x)^n)$ with $n\ge 3$, in Eq.~(\ref{eq:Psi_expansion}).
If we cut out a finite-volume sub-box realization from a large-box simulation, the super-box tidal force for the realization should include contributions from all the Fourier modes, 
around the box, including the modes comparable with the sub-box size that correspond to the higher-order terms we ignore. 
Such intermediate-wavelength modes are incompatible with 
the periodic boundary conditions employed in a SU simulation.
To study the limitation of the approximation, ideally, we can extend 
the method for super-survey density contrast ($\delta_{\rm b}$) in Fig.~2 of \citet{2014PhRvD..90j3530L} to the super-survey tidal force to test how the SU approach can accurately simulate the structure formation, even in the average sense. However, this is beyond the scope of this paper. 
Instead we use a larger-size box simulation to study how the results are sensitive to a chosen size of simulation box. 

Figure \ref{fig:gk-Lbox} compares the growth tidal response $G_K$ at $z=0$ and $2$ from the simulations with different box sizes 
of $L_{\rm box}=125$ and $500~\hiMpc$ (we will discuss in more detail the results of the growth response in the next section, so here let us focus on the comparison). 
The $L_{\rm box}=500~\hiMpc$ run employs $N_{\rm part}=512^3$ and $\lambda_{{\rm A},z}=0.01$ in four realizations,
and includes Fourier modes whose wavelengths are longer than
the box size of another simulation,  $L=125~\hiMpc$. 
The mass and spatial resolutions of the $L_{\rm box}=500~\hiMpc$ run are as same as the LR sets.
According to the resolution study discussed in Sec. \ref{sec:resolution}, 
the results from the $L_{\rm box}=500~\hiMpc$ 
should be reliable
up to $k=1.5~\hMpci$.
We confirm that 
the results for the different box sizes
are consistent with each other in the reliable range of $k$.
Thus we conclude that $L_{\rm box}=125~\hiMpc$ is large enough to measure the tidal response at $z\leq2$.
A larger-box simulation allows us to explore $G_K$ at smaller-$k$.
We find that $G_K$ from the $L_{\rm box}=500~\hiMpc$ run well agrees with the perturbation theory prediction even at $k\gtrsim0.01~\hMpci$.
Thus as long as a simulation box size is in the linear regime at an output redshift, we can use SU simulations of arbitrary box size to estimate the 
response function. This is the similar condition to that for a SU simulation of the super-survey density contrast ($\delta_{\rm b}$)
\citep{Li14}.

Hence hereafter we use the simulations of box size $L_{\rm box}=125~\hiMpc$ as our default choice.


}}

\section{Results}
\label{sec:res}

In this section we show the main results of this paper, i.e. the SU simulation results for the growth response function for the large-scale tidal force. In particular we present a validation of the response function computed from the TreePM code in the linear and 
nonlinear regimes by comparing the results with the perturbation theory results at large scales (small $k$) and with those from higher-resolution PM simulations.

\subsection{Monopole power spectrum: a sanity check of anisotropic SU simulations}
\label{sec:monopole}

Before going to the main results, we study the monopole power spectrum measured from the SU simulations. Although the large-scale tidal force causes an anisotropic distortion in the matter power spectrum depending on the alignments between the wavevector $\bk$ and the tidal tensor $K_{ij}$, the monopole spectrum should not be changed due to the traceless nature \citep{akitsu17}. Hence, the monopole spectrum measured from SU simulations should be free of the large-scale tidal effect to within the measurement errors, if SU simulations are properly implemented. Hence this gives a useful sanity check of SU simulations.

First we analytically check that the monopole spectrum measured in SU simulations should be free of the large-scale tidal effect. 
The monopole power spectrum is computed as
\begin{align}
    &\langle P_W(\bk_W, K)\rangle\nonumber\\
    &=\frac{1}{4\pi}\int {P}_W(\bk_W, K)\sin\theta_{\bk_W} d\theta_{\bk_W} d\phi_{\bk_W} \nonumber \\
    &\simeq\frac{1}{4\pi}\int P(k)\left[1+G_K(k)\hat k_i \hat k_j K_{ij}\right]
    \sin\theta_{\bk_W} d\theta_{\bk_W} d\phi_{\bk_W}\nonumber \\
    &=P(k)\left[1+\frac{G_K(k)}{4\pi}\int \hat k_i\hat k_j K_{ij}
    \sin\theta_{\bk_W} d\theta_{\bk_W} d\phi_{\bk_W}\right]=P(k), 
\end{align}
where \SM{$\theta_{\bk_W},\phi_{\bk_W}$ are the polar coordinate in the Fourier space and} we have used the traceless condition of $K_{ij}$ as
\begin{align}
    &\frac{1}{4\pi}\int \hat k_i\hat k_j K_{ij}\sin\theta_{\bk_W} d\theta_{\bk_W} d\phi_{\bk_W}\nonumber\\
    &=\frac{1}{4\pi}\int \hat k_i\hat k_j K_i\delta^{\rm K}_{ij}
    \sin\theta_{\bk_W} d\theta_{\bk_W} d\phi_{\bk_W}
    \simeq\frac{k_W^2}{3k^2}\sum_i K_i
    =0.
\end{align}

\begin{figure}
\begin{center}
\includegraphics[width=\hsize]{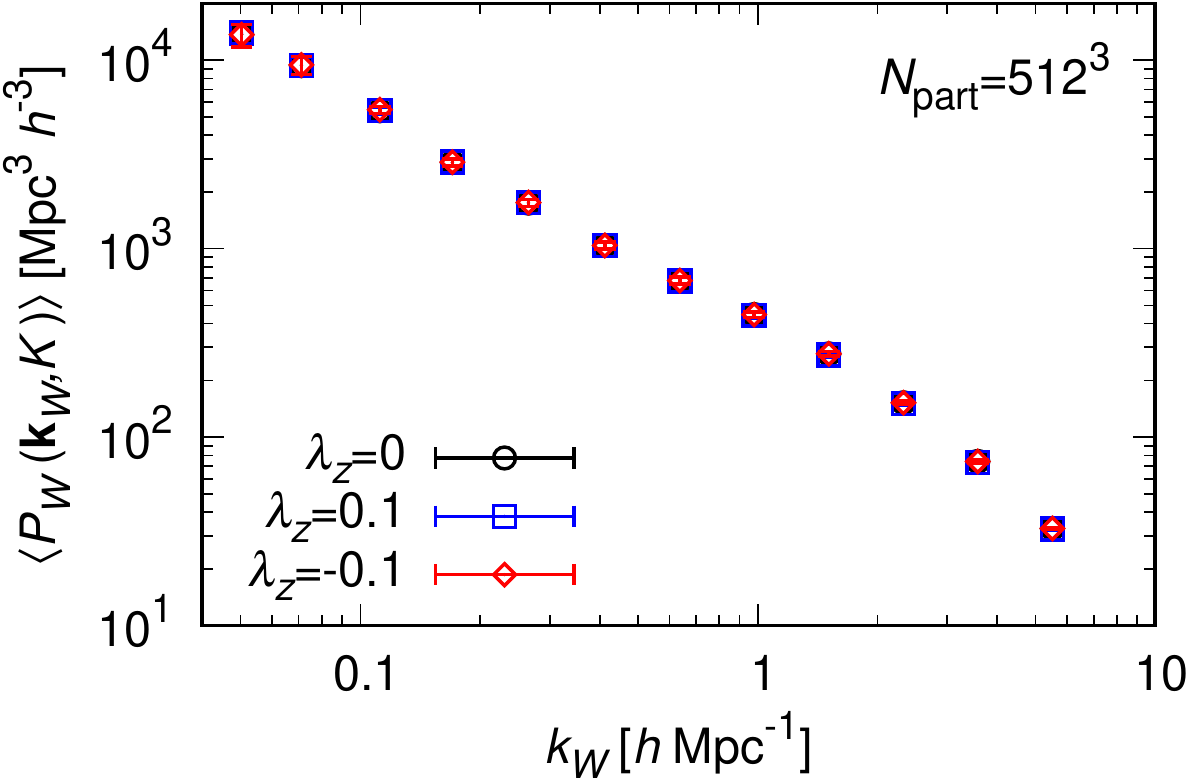}
\caption{The monopole power spectra at $z=0$, measured from SU simulations with 
$N_{\rm part}=512^3$ and  $\lambda_z=0.1$ (HR1-A) and $-0.1$ (HR1-B), respectively. The monopole power spectrum should be free of 
the large-scale tidal effect if the SU simulations are properly done, as discussed in Section~\ref{sec:monopole}. 
These results can be compared with the matter power spectrum measured from the isotropic background simulation (i.e. the standard 
simulation), with $N_{\rm part}=512^3$ and $\lambda_z=0$ (HR-C).
}
\label{fig:pk-tidal}
\end{center}
\end{figure}
Figure~\ref{fig:pk-tidal} clearly shows the monopole power \smrv{spectra} 
measured from SU simulations with
$N_{\rm part}=512^3$ and $\lambda_z=0.1$ (HR1-A), $-0.1$ (HR1-B) and $0$ (HR-C)
well agree with each other. 
%
\begin{figure*}
\begin{center}
\includegraphics[width=0.32\hsize,angle=0]{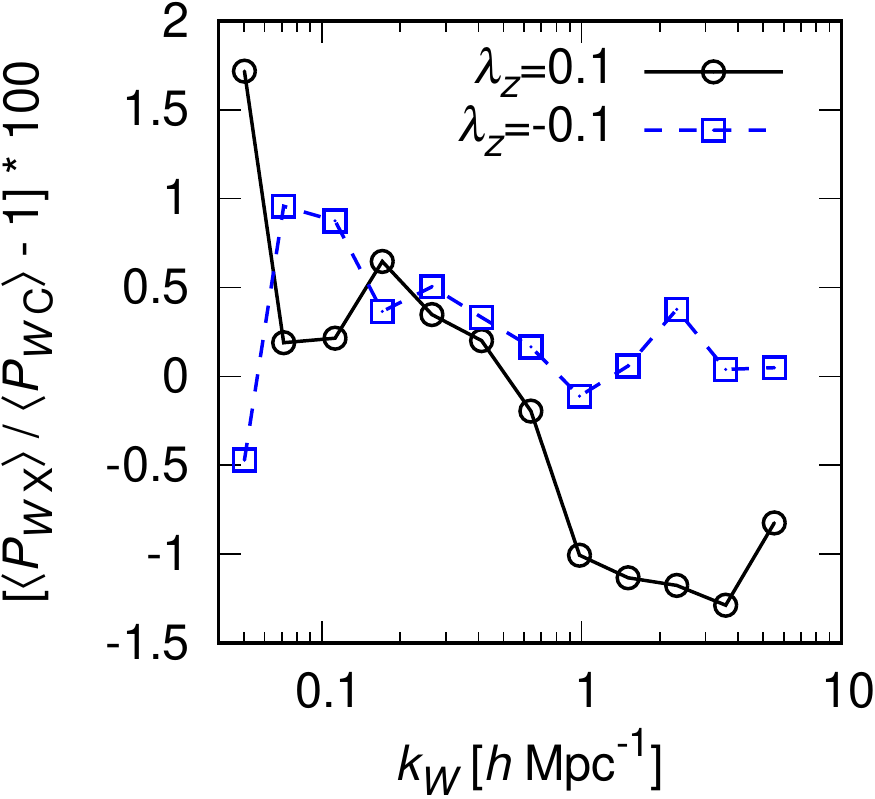}
\includegraphics[width=0.32\hsize,angle=0]{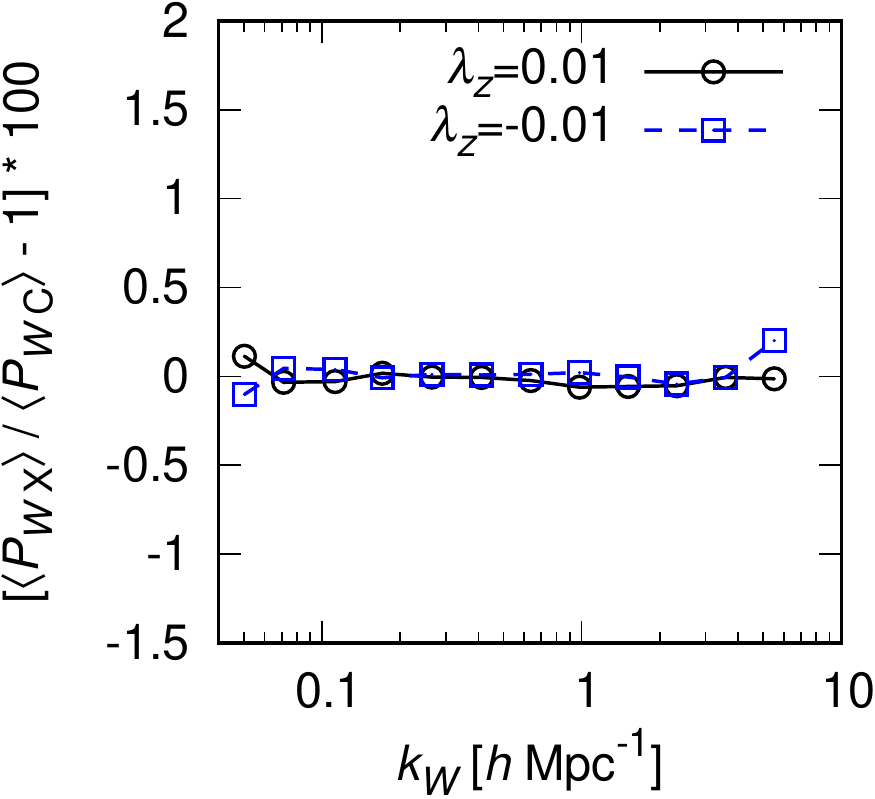}
\includegraphics[width=0.32\hsize,angle=0]{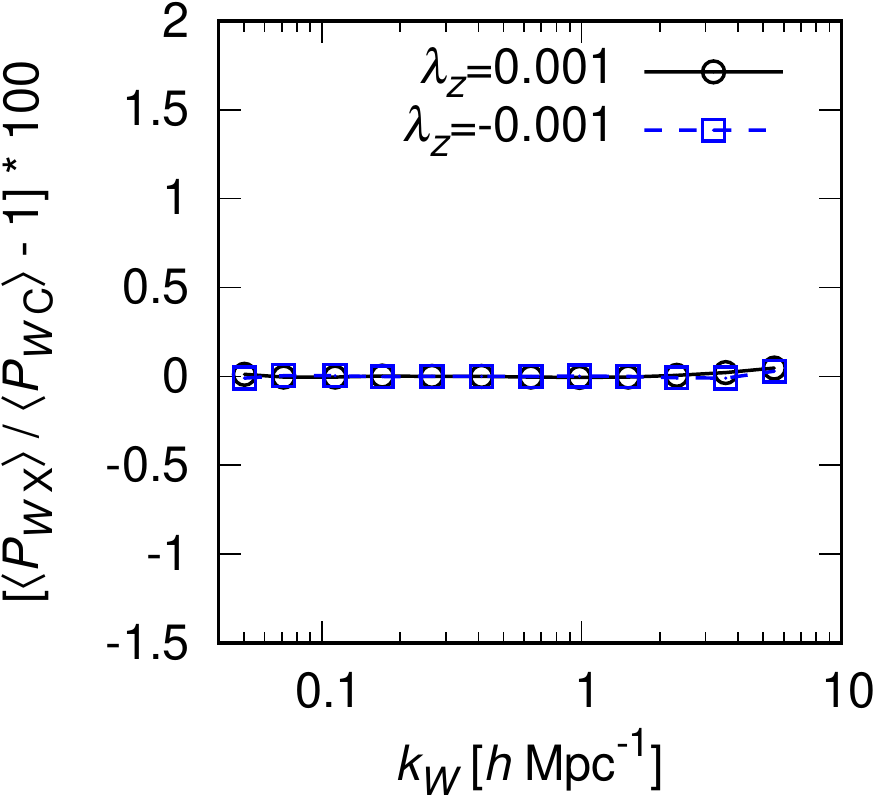}
\caption{Similar to the previous figure, but each panel shows the fractional difference of the monopole power spectra for the 
SU simulations relative to the power spectrum
in the isotropic universe simulation, denoted as $\langle P_{W \rm C}\rangle$. 
Here we show the results for the SU simulations with $\lambda_z=\pm0.1$ (HR1), $\pm0.01$ (HR2) and $\pm0.001$ (HR3) from left to right panels, respectively. The subscript ``X'' in $\langle P_{W \rm X}\rangle$ denotes the SU simulation runs with ``A'' or ``B'' in Table~\ref{tab:sims}.
}
\label{fig:pk-diff}
\end{center}
\end{figure*}
To see in more detail, we show fractional differences between  the power spectra of the isotropic and anisotropic runs in 
Figure~\ref{fig:pk-diff}.
All results are very close to zero better than a sub percent in the fractional amplitude for the cases of $|\lambda_z|=0.01$ and 0.001 over the range of $k$ we study. The results for $|\lambda|=0.1$ shows a percent-level deviation, implying a nonlinear contribution of $O(K^2)$, i.e. a violation of the linear response formula. 
Therefore the $|\lambda_z|$ values less than $0.01$ are safer to obtain the linear response function to $K_{ij}$, being not affected by the higher-order effect of $K_{ij}$.

\begin{figure}
\begin{center}
\includegraphics[width=\hsize]{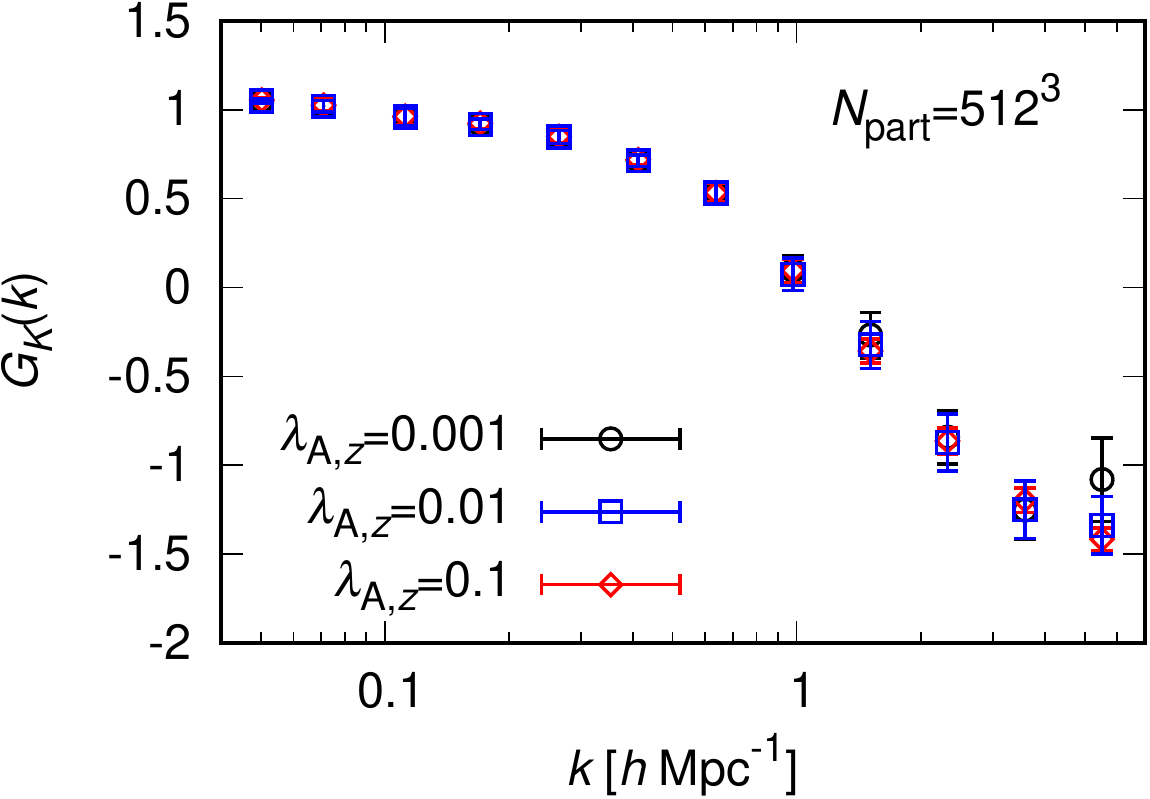}
\caption{The growth response, $G_K(k)$, at $z=0$, measured from paired SU simulations with 
$N_{\rm part}=512^3$ and $\lambda_{z}=\pm 0.001$ (HR3), $\pm 0.01$ (HR2) and $\pm 0.1$ (HR1).}
\label{fig:gk-tidal}
\end{center}
\end{figure}

\subsection{\mtrv{Growth response $G_K(k)$ in the linear and \smrv{nonlinear} regimes}}
\label{sec:growth_response}

In Figure~\ref{fig:gk-tidal} we study the growth response, $G_K(k)$, measured from the paired SU simulations with
$N_{\rm part}=512^3$ and $\lambda_{{\rm A},z}=0.001$ (HR3), $0.01$ (HR2) and $0.1$ (HR1), using the estimator Eq.~(\ref{eq:est_GK}).
All the results for $G_K$ remarkably well agree with each other 
over the range of scales we consider, to within the statistical error bars. 
One notable advantage is the use of paired SU simulations using the same initial seeds allow for an accurate calibration of 
the growth response, $G_K(k)$, even at very small $k$, where the sample variance is large.
Conversely, the results at small scales $k\gtrsim 1~h~{\rm Mpc}^{-1}$ appear to be relatively noisy.

\begin{figure}
\begin{center}
\includegraphics[width=\hsize]{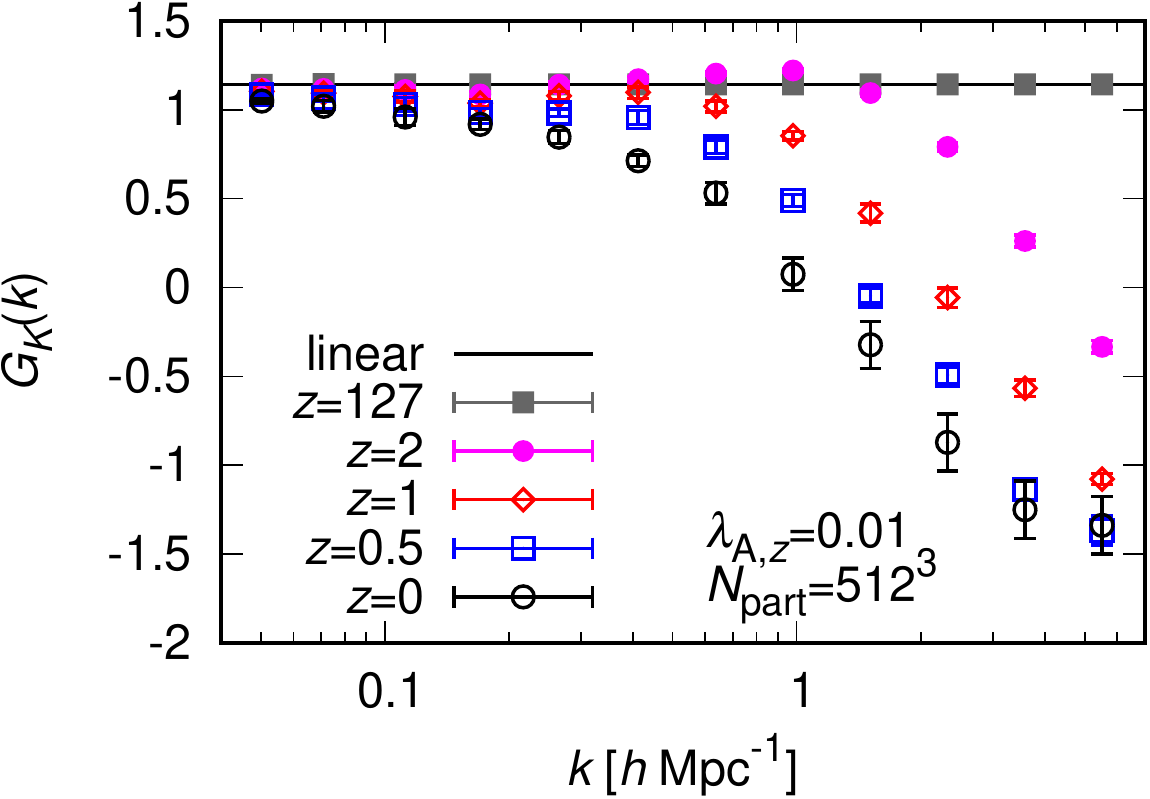}
\caption{Shown is how the growth response evolves with redshifts; the results measured from the SU simulations at outputs
$z=127, 2, 1, 0.5$ and $0$, respectively, are shown. 
We used the SU simulations 
with $N_{\rm part}=512^3$ and $\lambda_{z}=\pm 0.01$ (HR2).
The solid line shows the perturbation theory 
prediction of $G_K=8/7$ that should hold at the limit of $k\rightarrow 0$.}
\label{fig:gk-z}
\end{center}
\end{figure}
Figure~\ref{fig:gk-z} shows how the growth response, $G_K(k;z)$, evolves with redshift, measured 
from the simulation sets
of HR2-A, HR2-B and HR-C, i.e., $N_{\rm part}=512^3,\lambda_{{\rm A},z}=0.01$.
The solid line gives the perturbation theory prediction of 
$G_K=8/7$ \citep{akitsu17}, which should be valid at the limit of $k\rightarrow 0$. 
By construction of the initial conditions, the simulations at $z=127$ reproduces $G_K=8/7$ (see around Eq.~\ref{eq:pk-ic}).
As structure 
evolves, the simulation results start to deviate from the perturbation theory prediction due to the nonlinear mode coupling; the SU simulations lead to the smaller amplitudes in the growth response in the weakly nonlinear regime, and the response 
eventually turns to be
 negative at smaller scales \citep{schmidt18}.
Comparing the results at different redshifts also manifests that 
the simulation results match the perturbation theory prediction up to higher $k$ for higher redshifts, as expected.  
At $z=0$, the nonlinear effect is important already at $k\gtrsim 0.1~h{\rm Mpc}^{-1}$.
We have compared our results at $z=0,1$ with \citet{schmidt18}, and found that the agreements are fairly good.

\begin{figure}
\begin{center}
\includegraphics[width=\hsize]{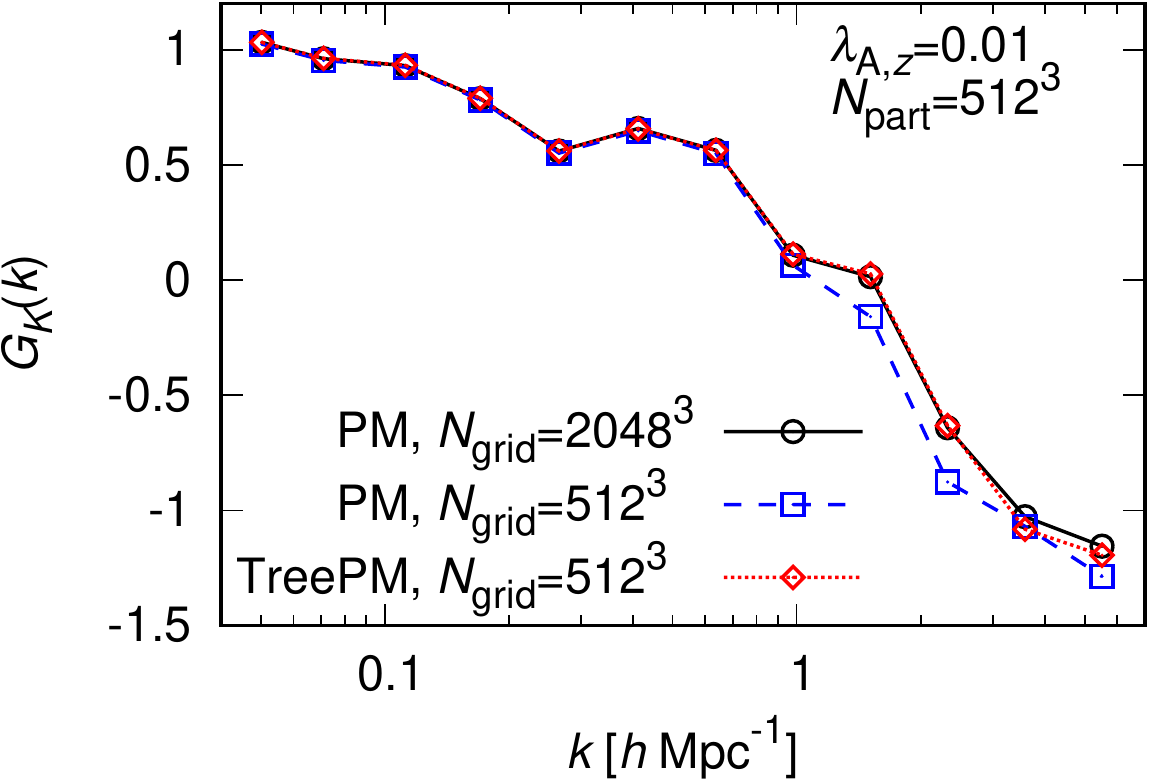}
\caption{A validation of the growth response measurement using 
the TreePM run with $N_{\rm grid}=512^3$ and $\lambda_z=\pm 0.01$
in our SU simulation implementation, compared with the high-resolution PM-only SU simulation with $N_{\rm grid}=2048^3$. 
For comparison we also show the result for the PM-only simulation with $N_{\rm grid}=512^3$.
}
\label{fig:gk-pm-tpm}
\end{center}
\end{figure}
Figure~\ref{fig:gk-pm-tpm} gives a validation of the TreePM based SU simulation, addressing
whether 
our Tree part computes gravity in the anisotropic SU simulation 
properly. 
As we stated above, our PM part works well because $G_K$ in the linear scales agrees with the 
perturbation theory
prediction of $G_K=8/7$ \smrv{(see also Figure \ref{fig:gk-Lbox} for the agreement between the results from the larger-box run and the perturbation theory prediction at smaller-$k$,  $k\gtrsim0.01~\hMpci$)}.
To test the implementation of the Tree part, we additionally perform two PM-only runs,  
and compare their results with the TreePM result 
for 
$G_K$.
One of the PM-only runs employ 
$N_{\rm grid}=512^3$ as in our default TreePM run. 
The difference between the two runs is whether the short range force is supplemented by the Tree force or not.
Then we use the higher-resolution PM\tnrv{-only} run with 
$N_{\rm grid}=2048^3$ 
to validate the TreePM run, 
by supplementing
the gravitational force on small separations with the refined grid resolution. 
We run the high-resolution PM runs assuming 
a set of the three initial conditions as in 
the sets of HR2-A, HR2-B and HR-C. 
We find that the PM run with $N_{\rm grid}=512^3$ underpredicts 
the $G_K$ amplitudes
by up to 40\% at $k>1~h~{\rm Mpc}^{-1}$ compared to the PM run with $N_{\rm grid}=2048^3$.
On the other hand, the TreePM run with $N_{\rm grid}=512^3$ agrees with the 
high-resolution PM run to within 5\% level 
even at $k>1~h~{\rm Mpc}^{-1}$.
This agreement is encouraging because it means
that the Tree part in our SU simulation properly solves gravity in structure formation down to 
small scales. 
The slight mismatch at the two largest $k$ bins between the TreePM and the PM-only run with the larger number of grids would be a sign that the latter starts to fail to resolve the force. Unlike the adaptive nature of the force resolution from a TreePM calculation, the fixed force resolution with the PM-only runs should eventually see a breakdown. Since the lower resolution PM-only run (with $N_\mathrm{grid}=512^3$) starts to deviate from the other two at around $k\simeq1\,h~\mathrm{Mpc}^{-1}$, the higher-resolution PM-only run, which has four times higher resolution per dimension ($N_\mathrm{grid}=2048^3$), is expected to perform poorly at $k\simeq4\,h~\mathrm{Mpc}^{-1}$. This is indeed about the scale at which the TreePM run and the higher-resolution PM-only run 
start
to deviate. Therefore, we conclude that our TreePM implementation works as expected compared to the PM-only runs.

\subsection{Total response $R_K(k)$ in the linear and nonlinear regimes}

\begin{figure}
\begin{center}
\includegraphics[width=\hsize]{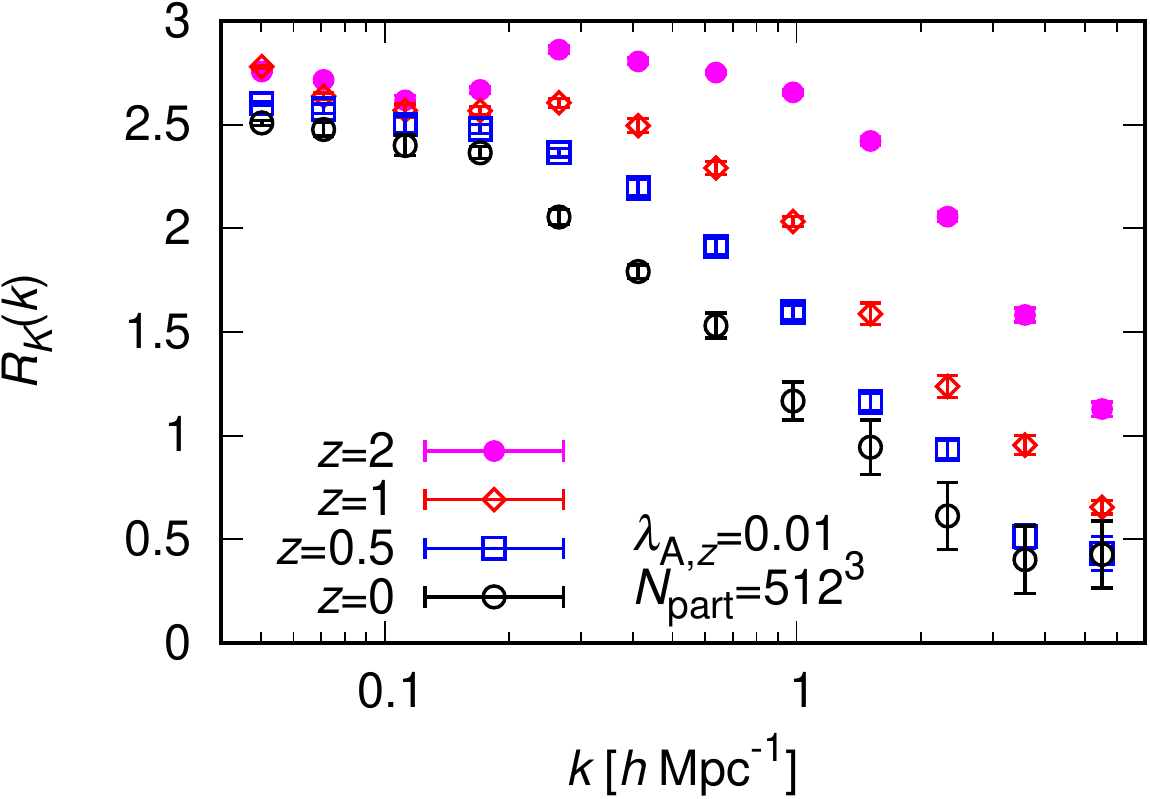}
\caption{The total response to the large-scale tidal force at different redshifts, which is given by a sum of the growth response and the dilation response: 
$R_K(k)=G_K(k)-\partial \ln P(k)/\partial \ln k$. We used the SU simulations with 
$N_{\rm part}=512^3$ and $\lambda_{z}=\pm 0.01$ (HR2) to evaluate the growth response, and employ the emulator {\sc Cosmic Emu} 
\citep{MiraTitan1,MiraTitan2} to evaluate the dilation response for the nonlinear power spectrum,
$\partial \ln P(k)/\partial \ln k$.
}
\label{fig:rk-z}
\end{center}
\end{figure}
Finally we study the full response that is given by a sum of the growth response and the dilation response: 
$R_K\equiv G_K-\partial \ln P(k)/\partial \ln k$ (see Eq.~\ref{eq:RK-GK}). 
We use the \smrv{emulator} {\sc Cosmic Emu} \citep{MiraTitan1,MiraTitan2} to 
evaluate  the dilation term $\partial\ln P/\partial \ln k$ that should be valid up to the nonlinear scales.
The figure clearly shows that the total response has a $k$-dependence, and has smaller amplitudes at very large \smrv{$k$,}
indicating 
an asymptotic behavior of 
$R_K\rightarrow 0$ at very large $k$. However, up to $k\simeq 6~h~{\rm Mpc}^{-1}$, the total response has a non-vanishing value, 
$R_K>0$, meaning that the large-scale tidal force affects structure formation even at such small scales. 
Note that the baryonic acoustic oscillation features at $k\simeq 0.1~h~{\rm Mpc}^{-1}$ are not seen due to our sparse $k$-binning.

\section{Conclusion and prospects}
\label{sec:conc}

In this paper
we have developed a TreePM cosmological $N$-body simulation code to simulate structure formation in a finite volume including the effect of super-box tidal force ($K_{ij}$) on sub-box modes. To do this, we have presented a formulation to include the effect of super-box tidal force into the anisotropic background expansion, which can be done by introducing the anisotropic scale factor, $a_{Wi}(t)$ -- so-called anisotropic separate universe (SU) simulation. We modified the public $N$-body code, {\sc Gadget-2}, to implement  
SU simulations for the $\Lambda$CDM cosmology. 
Extending the ``isotropic'' SU simulation technique for the super-box density contrast $\delta_{\rm b}$, which has been developed by many groups \citep{Li14,Wagner15a,2016JCAP...09..007B}, we modified both parts of PM force and Tree force to include the super-box tidal effects on structure formation on both large- and small-scales up to the deeply nonlinear regime. The modification of PM force is straightforward. However, the modifications of Tree force and the split factor that divides the forces into the PM and Tree forces are not trivial, and our treatment is slightly different from the previous work \citep{schmidt18}. We have tested and
validated our method by comparing the simulation results with the perturbation theory predictions and with the result from high-resolution PM code (Figures~\ref{fig:pk-tidal} -- \ref{fig:gk-z} and \ref{fig:gk-pm-tpm}). 
\smrv{We also study the impact of the periodic boundary condition on anisotropic SU simulations in Figure~\ref{fig:gk-Lbox}.}

We used the SU simulations to calibrate the ``response'' function of \tnrv{the} matter power spectrum that describes how the super-box tidal force affects the matter power spectrum as a function of wavenumber and redshift, for a given global cosmological model, the standard adiabatic $\Lambda$CDM model considered in this paper. 
With the aid of the TreePM SU simulations, we studied the response function over a wide range of scales from the linear to deeply nonlinear regimes, 
$0.05<k/[h~{\rm Mpc}^{-1}]\lesssim 6$. We showed that the response function has a characteristic $k$-dependence, and the large-scale tidal force affects structure formation over all the scales we have considered, i.e. down to the deeply nonlinear regime 
(Figures~\ref{fig:gk-z} and \ref{fig:rk-z}). 

As discussed in \citet{2014PhRvD..90j3530L} \citep[also see][]{akitsu17}, the leading-order response of large-scale fields to small-scale structure formation arises from the super-survey density contrast ($\delta_{\rm b}$) and the super-survey tidal tensor ($K_{ij}$) that are both from the second derivative tensor of the long-wavelength gravitational potential, 
reflecting
the nature of the second-derivative differential equations in Newtonian (Einstein) gravity. Hence a combination of isotropic and anisotropic SU simulations allows one to simulate structure formation in a finite volume, with periodic boundary conditions, not only including the effects of both super-box modes, but also keeping high resolution to accurately simulate nonlinear structure formation. As long as the density fluctuations of a  simulation box size are in the linear regime, SU simulations are valid.
In particular, the use of TreePM code based SU simulations allows one to simulate nonlinear structure formation including the effects of super-box modes on properties of halos where galaxies and galaxy clusters form, which therefore have direct connections to large-scale structure observables. 

There are many applications of anisotropic SU simulations to large-scale structure cosmology. Such applications include an accurate calibration of the tidal bias of halos and galaxies \citep{baldauf12,2018JCAP...01..010M}, the effects on redshift-space clustering of galaxies \citep{akitsu17,akitsu18,Akitsu19}, a calibration of the covariance matrix for the redshift-space power spectrum and weak lensing \citep{akitsu17,2018JCAP...02..022L,2018JCAP...06..015B,2019arXiv191002914W}, the intrinsic alignments of halo/galaxy shapes \citep{2012PhRvD..86h3513S,2015JCAP...10..032S,2016PhRvD..94l3507C,Okumura18,2020MNRAS.493L.124O,2020ApJ...891L..42T} (also see Kurita, Takada et al. in preparation), cosmology with optical clusters \citep{Osato18,2020arXiv200203867S}, the effects on galaxy formation in hydrodynamical simulations \citep{Barreira19}, the effects on cosmic reionization and physics of intergalactic medium \citep{2020arXiv200202467D}, and so on. These are all interesting and worth to explore, and this is a list for our future work.

While this paper is under completion,  the preprint \citep{2020arXiv200306427S} was put forward  in arXiv. Our work is based on a similar motivation, but independent and is not affected by their work.

\section*{Acknowledgements}

\smrv{We would like to thank the referee of this paper Vincent~Desjacques for useful comments.}
We would like to appreciate Volker~Springel and Andreas S. Schmidt for making their codes publicly available.
We would like to thank Kazuyuki~Akitsu and Yin~Li for useful discussion.
We would like to acknowledge Yosuke~Kobayashi for providing us with an analysis code of simulation data.
All simulations were carried out on Cray XC50 at Center for Computational Astrophysics, National Astronomical Observatory of Japan.
This work was supported in part by World Premier International Research Center Initiative (WPI Initiative), MEXT, Japan, JSPS KAKENHI Grant Numbers JP15H03654, JP15H05887, JP15H05893, JP15H05896, JP15K21733, JP17K14273 and JP19H00677, and Japan Science and Technology Agency CREST JPMHCR1414.

\bibliographystyle{mnras}
\bibliography{lssref}

\bsp
\label{lastpage}
\end{document}